\documentclass[amsmath,superscriptaddress,showpacs,aps,prl,twocolumn]{revtex4-1}
\usepackage{bm}
\usepackage{graphics}
\usepackage{graphicx}
\usepackage{amsthm}
\usepackage{amsmath}
\usepackage{bbding}

\usepackage{marvosym}
\usepackage{wasysym}
\usepackage{amssymb}
\usepackage{enumerate}
\usepackage{xfrac}
\usepackage[dvips]{epsfig}
\usepackage[caption=false]{subfig}  
\usepackage[normalem]{ulem}
\usepackage{color}
\usepackage{txfonts}
\usepackage{ulem}

\begin{document}
\title{Momentum Space Entanglement Spectrum of Bosons and Fermions with Interactions}
\author{Rex Lundgren}
\email{rexlund@physics.utexas.edu}
\author{Jonathan Blair}
\affiliation{Department of Physics, The University of Texas at Austin, Austin, TX 78712, USA}
\author{Martin Greiter}
\affiliation{Insitute for Theoritical Physics, Univesity of W\"{u}rzburg, D-97074 W\"urzburg, Germany}
\author{Andreas L\"{a}uchli}
\affiliation{Institut f\"{u}r Theoretische Physik, Universit\"{a}t Innsbruck, Technikerstra\ss e 25, A-6020 Innsbruck, Austria}
\author{Gregory A. Fiete}
\affiliation{Department of Physics, The University of Texas at Austin, Austin, TX 78712, USA}
\author{Ronny Thomale}
\affiliation{Insitute for Theoritical Physics, Univesity of W\"{u}rzburg, D-97074 W\"urzburg, Germany}
\begin{abstract}
We study the momentum space entanglement spectra of bosonic and fermionic formulations of the spin-1/2 $XXZ$ chain with analytical methods and exact diagonalization. We investigate the behavior of the entanglement gaps, present in both formulations, across quantum phase transitions in the $XXZ$ chain. In both cases, finite size scaling reveals that the entanglement gap closure does not occur at the physical transition points. For bosons, we find that the entanglement gap observed in [Thomale {\it et al.}, Phys.~Rev.~Lett.~{\bf 105}, 116805 (2010)] depends on the scaling dimension of the conformal field theory as varied by the $XXZ$ anisotropy. For fermions, the infinite entanglement gap present at the $XX$ point persists well past the phase transition at the Heisenberg point. We elaborate on how these shifted transition points in the entanglement spectra may support the numerical study of phase transitions in the momentum space density matrix renormalization group.

\end{abstract}


\maketitle

{\bf \em Introduction} -- Quantum information ideas applied to condensed matter systems have revealed novel insights into exotic phases of matter \cite{RevModPhys.80.517}. Quantitatively, quantum information between two regions, $A$ and $B$, can be characterized by the groundstate reduced density matrix of $A$, $\rho_A$, and analogously $B$, $\rho_B$. For example, the entanglement entropy (EE) is given by $\mathrm{Tr}(\rho_A \mathrm{ln}\rho_A)=\mathrm{Tr}(\rho_B \mathrm{ln}\rho_B)$. The entanglement spectrum (ES) \cite{Li:prl08} (defined as the set of eigenvalues of a fictitious entanglement Hamiltonian, $H_e$ , with $\rho_A$ written as $e^{-H_e}$) is a useful tool in understanding topological states of matter and strongly correlated systems, including fractional quantum Hall (FQH) systems \cite{Li:prl08,Thomale_AC:prl10,PhysRevB.84.045127,Lauchli:prl10,PhysRevB.88.155307,PhysRevLett.108.256806,PhysRevB.86.165314,Dubail:prb12,PhysRevB.88.075313}, quantum spin chains \cite{PhysRevLett.105.116805,2012JSMTE..08..011A,Pollmann:prb10,PhysRevLett.109.237208,PhysRevA.87.012302,Franchini:2010kq,PhysRevB.87.235107} and ladders \cite{Poilblanc:prl10,Lauchli:prb12,Lundgren,Fradkin_Ladder,Cirac:prb11,PhysRevB.88.245137,Tanaka:pra12,PhysRevLett.108.227201,2012PhRvB..86i4412Y}, topological insulators \cite{Kargarian:prb10,PhysRevLett.104.130502,PhysRevB.82.241102,PhysRevB.87.035119,PhysRevLett.103.261601,PhysRevB.84.195103}, symmetry broken phases \cite{PhysRevB.88.144426,2013PhRvL.110z0403A}, and other systems in one \cite{2013arXiv1303.0741L,PhysRevB.88.125142,Deng:prb11,Turner:prb11,Pollmann:njp10} and two \cite{PhysRevB.87.045115,PhysRevB.87.155120,Lou:prb11,PhysRevLett.107.157001,PhysRevLett.105.080501,PhysRevB.87.241103,PhysRevB.88.075123,PhysRevB.88.115114,PhysRevB.88.195102,PhysRevB.87.035141,PhysRevX.1.021014,2013NJPh...15e3017S,2014arXiv1402.0503T} spatial dimensions.  These studies predominantly focused on real / orbital space entanglement. For many gapped systems, the energy spectrum of the edge states and ES are equivalent. This was proven by  X.L. Qi {\it et al.} \cite{Qi:prl12} and elaborated on in Refs. \cite{PhysRevB.84.205136,Dubail_proof:prb12,PhysRevB.86.045117, PhysRevB.88.245137}. There is no universal understanding of systems with a gapless bulk, where long range correlations are present \cite{Calabrese:pra08}.

The ES in momentum space has been explored in quantum spin chains \cite{PhysRevLett.105.116805} and ladders \cite{Lundgren}. A momentum partition is natural and physically relevant, as the low-energy formulation of one-dimensional systems involves the splitting of particles into left and right movers \cite{Gogolin:book}. A deeper understanding of the momentum space ES could help identify the most fruitful applications of momentum space density matrix renormalization group (DMRG) algorithms ~\cite{PhysRevB.53.R10445,RevModPhys.77.259,hallberg,legeza}. Gapless spin chains are one promising candidate for momentum space DMRG. For example, for chains with higher symmetry groups characterizing the parameters of the critical theories is a challenge for real space DMRG and motivates different formulations~\cite{sun}. The momentum space ES is also useful in characterizing disordered fermionic systems \cite{disorder_fermi_MS,2014arXiv1403.6129M,2014arXiv1403.2599A} as well as in quantum field theories, where large momenta were traced over\cite{PhysRevD.86.045014}.

The momentum space ES of the spin-1/2 Heisenberg model exhibits a fingerprint of the underlying conformal field theory (CFT) in the counting of the entanglement levels and a large entanglement gap (EG), a notion first observed in the conformal limit construction of FQH entanglement spectra~\cite{Thomale_AC:prl10}: The counting of the entanglement levels below the EG in the spin chain relates the U(1) boson counting in the gapless sine-Gordon regime to the U(1) edge of the bosonic Laughlin state. The spin chain EG becomes infinite at the Haldane-Shastry (HS) \cite{PhysRevLett.60.635,PhysRevLett.60.639} point, whose Fourier transformed wave function yields the same weights of monomials as the Laughlin state. This fits into a more general connection of certain critical quantum spin chains and FQH states \cite{Greiter,PhysRevB.85.195149}.

As recently pointed out, the ES may provide a useful indication of distinct phases, but not phase transitions \cite{2013arXiv1311.2946C,2013arXiv1312.0619B}. In particular, Ref.~\cite{2013arXiv1311.2946C} highlighted that phase transitions can occur in the real-space entanglement Hamiltonian even though the physical ground state remains the same. (Ref.~\cite{Thomale_AC:prl10} noted early on that transitions according to an EG closure appear shifted as compared to the physical system.) Physically, this is because the properties of a system are determined by $H_e$ at finite temperature. Furthermore, Ref.~\cite{2013arXiv1312.0619B} stressed that non-analyticities of the ES can be connected to symmetries in real space, and are not always linked to phase transitions. In view of numerical techniques such as DMRG, whose performance is directly tied to the ES, we wish to convey that the persistence of EGs beyond physical transitions might improve the performance of DMRG algorithms. If the entanglement weight below the gap still provides an effective representation of entanglement contained in the state, the presence of an EG promises a reasonably constant performance of the numerical algorithm as one sweeps over the physical phase transition.

In this Letter, we explore the ES in momentum space for fermions and bosons in the $XXZ$ spin-$1/2$ chain. We find in both cases the ES fails to capture features of physical phase transitions. For bosons, we find the EG seen by Thomale {\it et. al} \cite{PhysRevLett.105.116805} is not always observed for CFTs of the same central charge, $c$, but different scaling dimensions. We also find that the bosonic momentum space ES at the Heisenberg point is flat, and despite similarity to the FQH Laughlin state, lacks topological entanglement entropy (TEE) \cite{PhysRevLett.96.110405,PhysRevLett.96.110404,PhysRevLett.98.060401}. For fermions, we observe from finite size scaling that the EG does not capture the phase transition present in the $XXZ$ spin-1/2 chain. We argue that due to their deviation from the physical phase transitions, the properties of the entanglement spectral flow might prove {\em useful} for entanglement-based numerical applications.

\begin{figure}
\begin{center}
\includegraphics[width=1\linewidth]{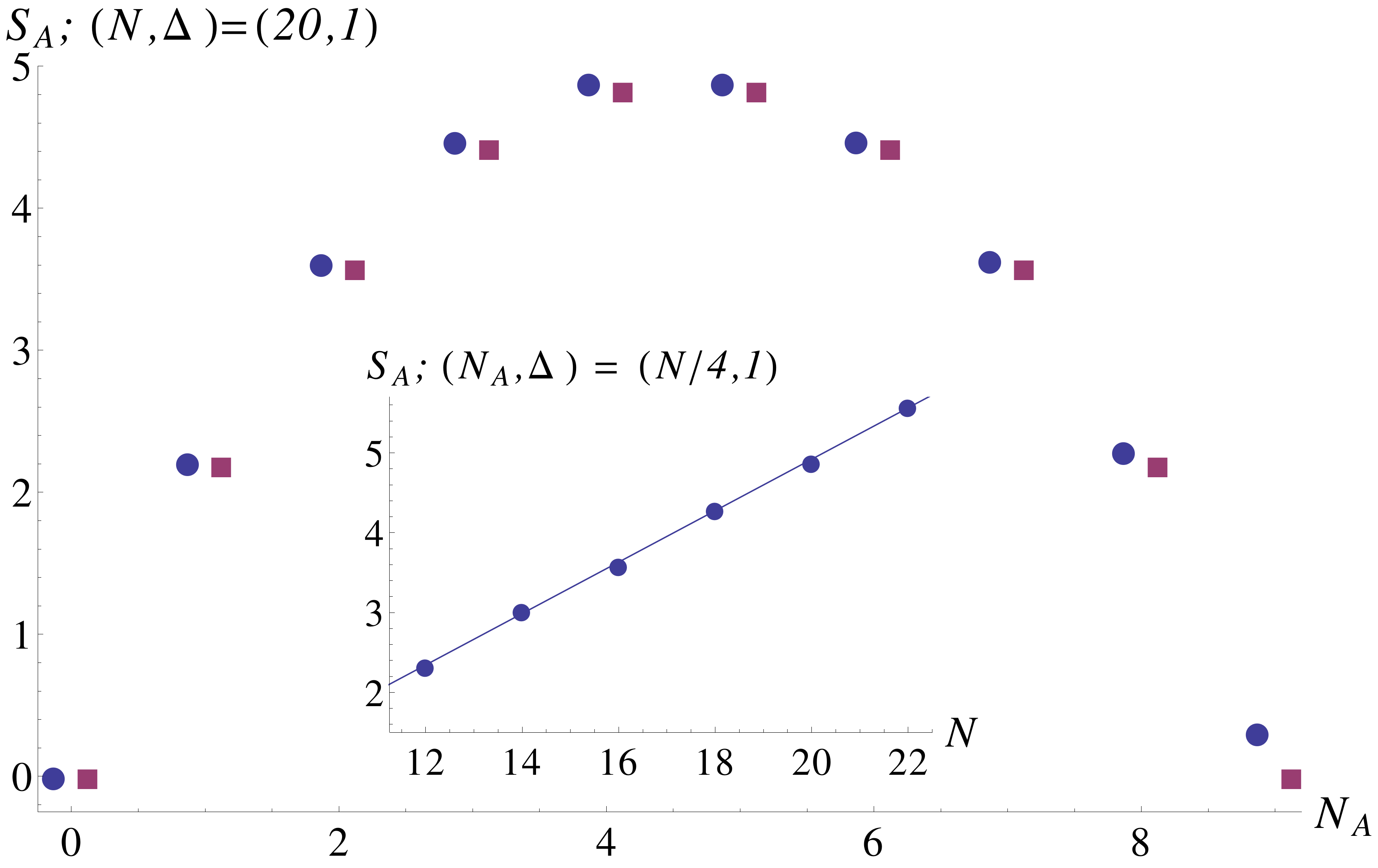}
\end{center}
\caption{(color online) EE of bosons, $S_A$, versus $N_A$ at $\Delta=1$ for $20$ sites (blue circle). EE assuming a flat ES (violet square). Inset: $S_A(N/4)$ versus $N$ ($N_A=N/4-1/2$ if $N/2$ is odd). $S_A$ is fit well by a linear equation, as expected for large $N$. \label{Flat}
}
\end{figure}

{\bf \em Model and Details of Partition} -- We investigate the $XXZ$ spin-1/2 chain, represented by hardcore bosons and by Jordan-Wigner fermions. The Hamiltonian with nearest and next-nearest neighbor interactions is given by
\begin{equation}
\label{Ham}
H=\sum_{n=1}^2\sum_{i=1}^N J_n(S^x_i S^x_{i+n}+S^y_i S^y_{i+n}+\Delta S^z_iS^z_{i+n}),
\end{equation}
with periodic boundary conditions (PBC) and length $N$. The transformation from spin operators to bosons is given by $S_i^+=b_i^{\dagger}$, $S_i^z=(b_i^{\dagger}b_i-\frac{1}{2})$, with an added hard-core term that prevents double occupancy. The hard-core term is an important source of entanglement in momentum space for bosons \cite{Supplemental} and leads to differences in the ES between bosonic and fermonic formulations. Most literature has focused on the similarity of physical properties of bosons and fermions in one dimension \cite{Giamarchi:book}; our work highlights a difference between bosons and fermions. The transformation to fermions is given by the Jordan-Wigner transform, $S_i^+=c_i^{\dagger}\prod\limits_{j=1}^{i-1}(1-2 c_{j}^{\dagger}c_j)$ and $S_i^z=(c_i^{\dagger}c_i-\frac{1}{2})$. The phase diagram of Eq.~\eqref{Ham} is well studied \cite{Giamarchi:book,PhysRevB.86.094417}.  We focus on the regime $0\leq \Delta<3$ and $J_2=0$ for the main part of the manuscript.  For $0\leq \Delta \leq 1$ the model is in a gapless phase with $c=1$. The fermions are free at $\Delta=0$ and acquire interactions of increasing strength with increasing $\Delta$; the bosons are strongly interacting throughout due to the hard-core term. $\Delta=1$ is the $SU(2)$ symmetric Heisenberg point. For $\Delta>1$, the model is in the gapped Ising phase. To set up a momentum space partition of the Hilbert space, we Fourier transform the bosons (fermions) as $(b,c)_j=\frac{1}{\sqrt{N}}\sum\limits_{m=-\frac{N}{2}+1}^{\frac{N}{2}}e^{imj}(b,c)_m$, where $m$ is the crystal momentum. Momentum basis states are labeled by occupation number, $n_m$, and crystal momentum, $\frac{2\pi m}{N}$, $m \in \{-\frac{N}{2}+1,\ldots,\frac{N}{2}\}$. The ground state of Eq.~\eqref{Ham} has $S_{tot}^{z}=\sum\limits_{i=1}^N S_i^z=0$ for the range of parameters we consider. Thus, the number of bosons and fermions, $N_p$, is $\frac{N}{2}$. For fermions, we consider system sizes of $4n+2$, $n \in \mathbb{N}$, to avoid a degenerate Fermi sea at $\Delta=0$. 

\begin{figure}
\begin{center}
 \includegraphics[width=1\linewidth]{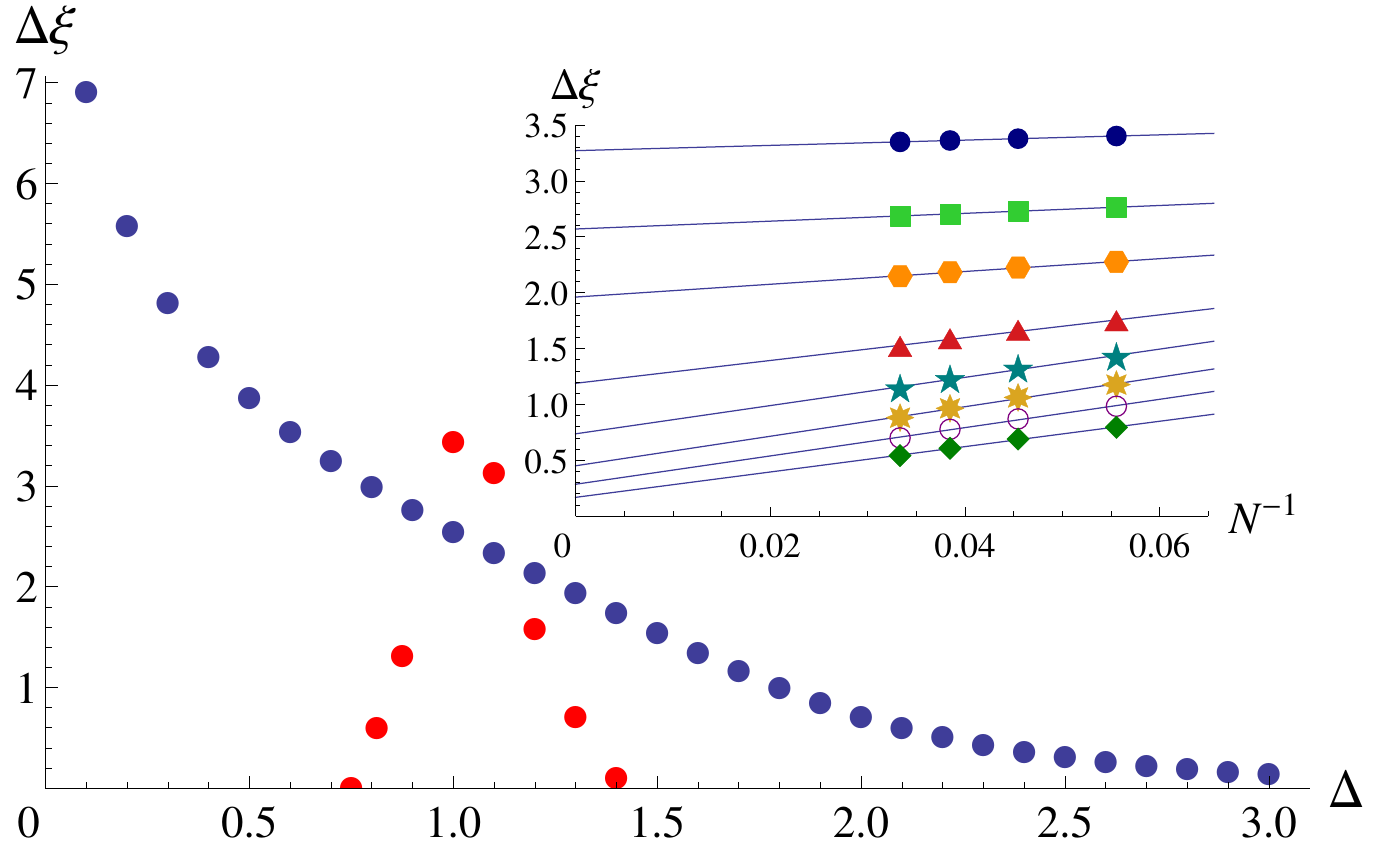}
\end{center}
\caption{(color online)  The TD projection $[$from finite size scaling with $N_A=N_p/2$ ($N_A=\frac{N_p-1}{2}$ if $N_p$ is odd)$]$ of the EG, $\Delta \xi$, for fermions (blue circle) and bosons (red circle) versus $\Delta$. Inset: Finite size scaling of the EG for fermions versus $\Delta$. All data fit the form $\Delta\xi=\alpha +\frac{\beta}{N}$, where $\alpha,\beta>0$. Parameters: $\Delta=.7$ (navy blue disk),  $\Delta=1$ (green square), $\Delta=1.3$ (orange hexagon), $\Delta=1.6$ (red triangle), $\Delta=2.0$ (teal star), $\Delta=2.3$ (yellow eight-point star), $\Delta=2.6$ (purple empty circle),\text{ and } $\Delta=3.0$ (green diamond).}
\label{GAP}
\end{figure}

\begin{figure*}
\subfloat[][$\Delta=1$]{
 \includegraphics[width=.33\linewidth]{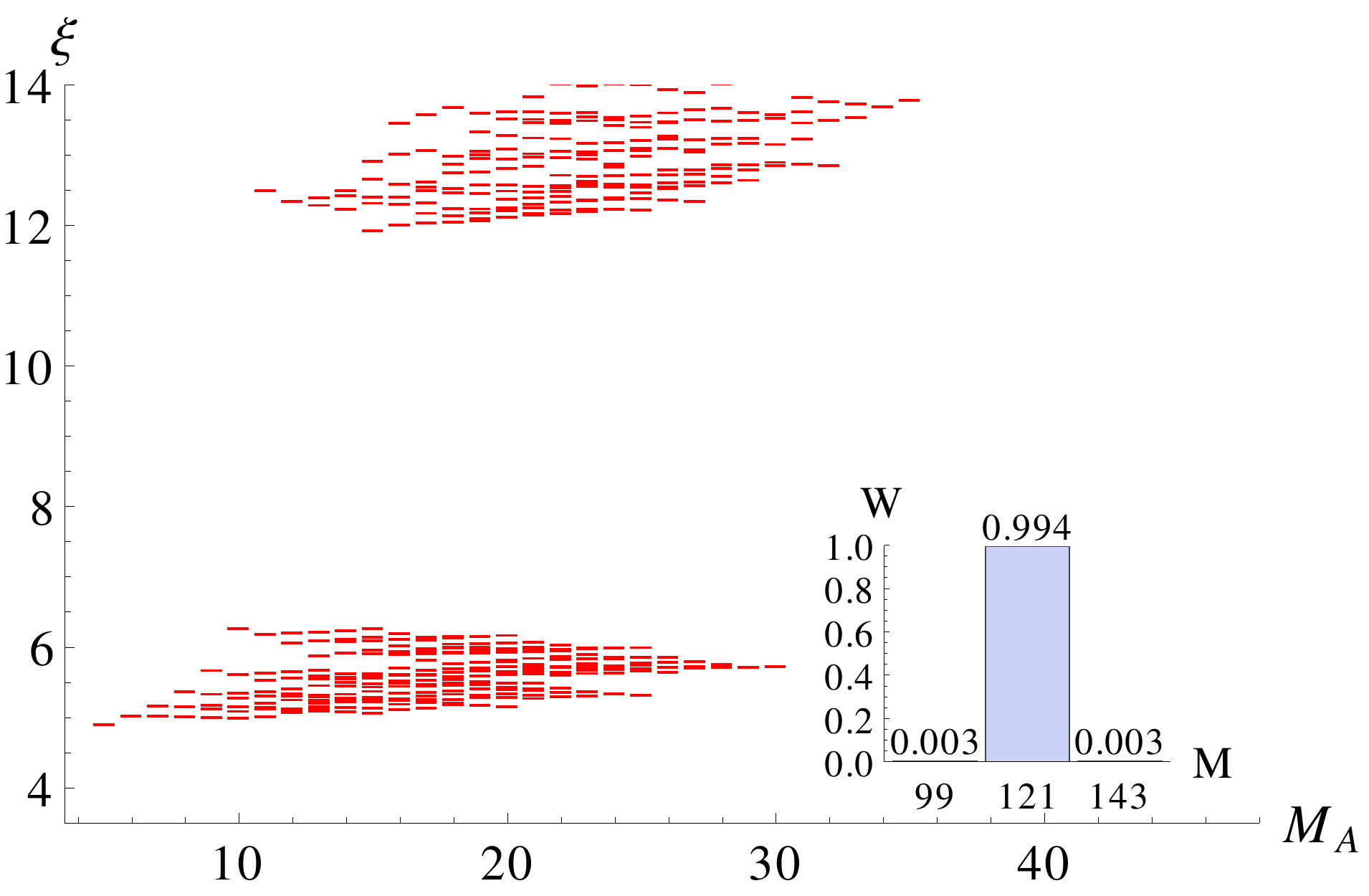}
}
\subfloat[][$\Delta=\frac{1}{2}$\label{bosons_one_half}]{
 \includegraphics[width=.33\linewidth]{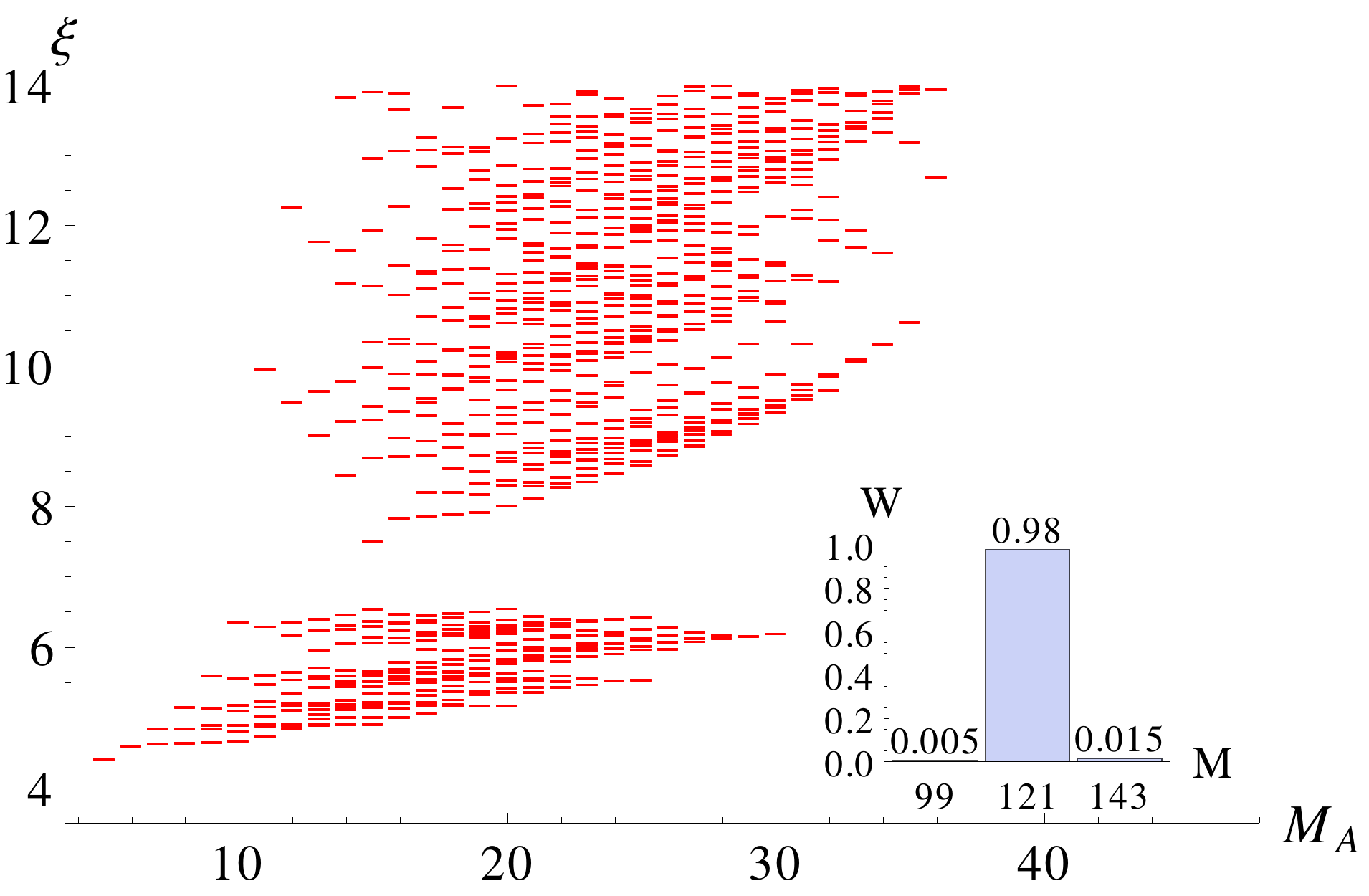}
}
\subfloat[][$\Delta=0$]{
 \includegraphics[width=.33\linewidth]{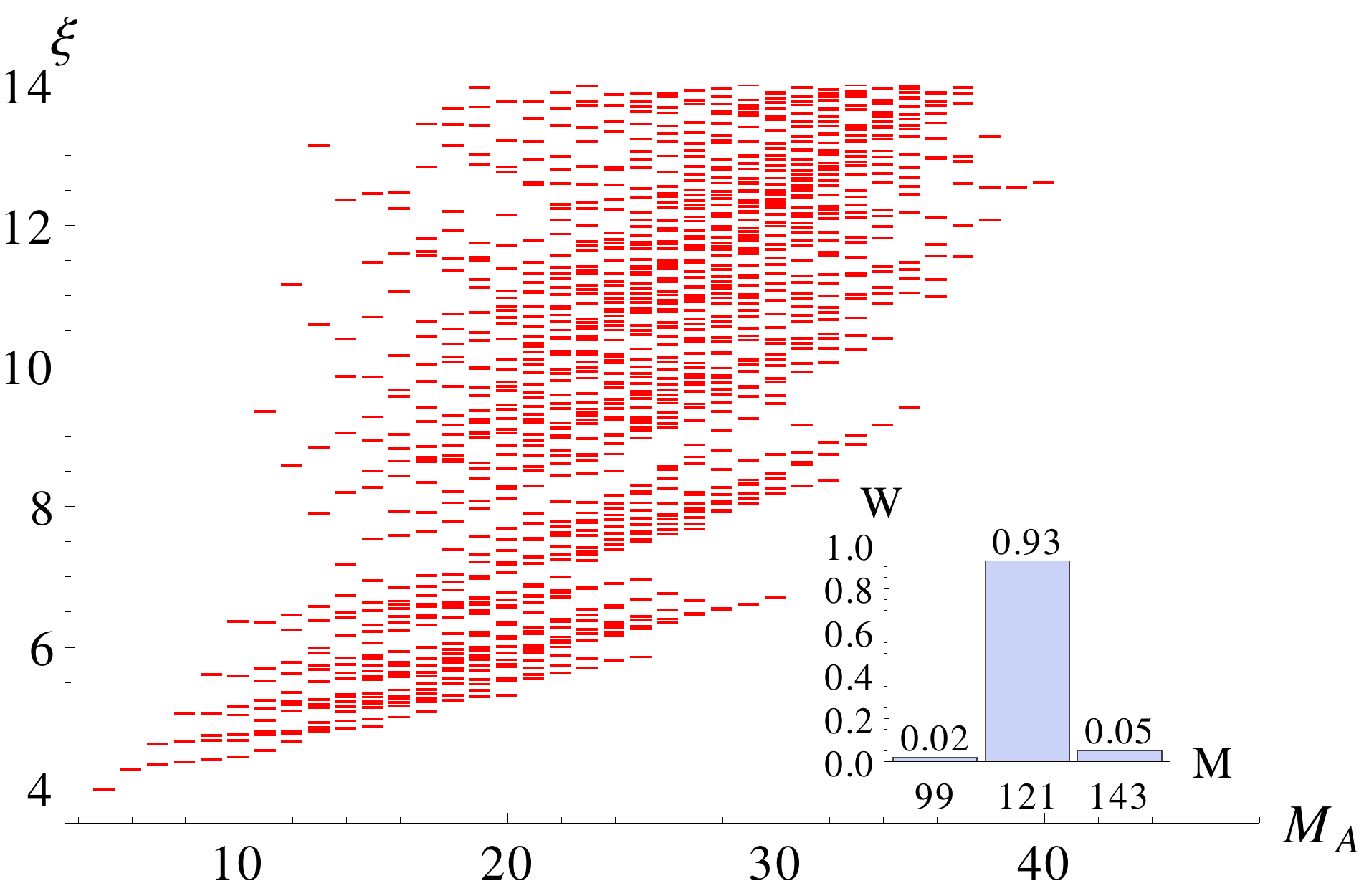}
}
\caption{(color online) Bosonic ES for representive values of $\Delta$, with $N=22$ and $N_A=5$. The entanglement eigenvalues $\xi$ are plotted versus $M_A$. Inset: Fraction of weight of the ground state, $W$, versus $M$. The ground state primarily resides in the $M=N^2/4$ sector.}
\label{Boson_ES_22}
\end{figure*}

After the ground state is obtained via exact diagonalization in the momentum occupation basis, we partition the system into two regions, $A$ and $B$, by dividing the momentum occupation basis at $\Gamma$, i.e. $m \in A=\{m~\vert~m > 0 \}$ and $m \in B =\{m~\vert~m~\le 0 \}$ and fixing the particle number in each region. We form the density matrix and then trace out the degrees of freedom of $B$. This yields $\rho_A$ and $H_e$. The total momentum, $M=\sum_m n_m m$, is only conserved approximately, and maps to the exact crystal momentum quantum number via $M_c=M\mod N$. Still, the total momentum can be a useful approximate quantum number when a large percentage of the weight is located in one sector as seen for the bosonic ground states near $\Delta=1$ in Fig.~\ref{Boson_ES_22}. (At the HS point, the total bosonic momentum becomes an exact quantum number \cite{PhysRevLett.105.116805}.) In such a case, we partition with respect to both number of particles and total momentum with the constraints $N_A+N_B=N_p$ and $M_A+M_B=M$. For a wave-function with strongly distributed weight in several momentum sectors (which happens for fermions), the total momentum is not a valid quantum number, and several momentum sectors will be mixed. In this case we partition the system with respect to the number of particles and the total crystal momentum of $A$ and $B$, i.e. $M_{A,c}+M_{B,c}=M_{c}$.  One can visualize the momentum space cut as a tracing out one-half of two coupled {\em chiral} one-dimensional systems.  Due to the numerical limitations of exact diagonalization, we consider systems sizes up to $N=22$ for bosons and $N=30$ for fermions.

{\bf \em Revisiting The Heisenberg Point} -- Ref.~\cite{PhysRevLett.105.116805} hinted that the ES for bosons below the EG is flat. We now show numerical evidence for this and analyze the consequences of a flat ES. We find the average of the levels at each $M_A$ below the EG approach the same constant value in the thermodynamic (TD) limit approximately equal to the natural log of the number of levels below the EG, $N_g=\frac{(N_p-1)!}{(N_p-1-N_A)!(N_A)!}$. This numerically demonstrates the ES is flat, so we can investigate how the EE scales with $N$. Neglecting the levels above the EG (this approximation is exact at the HS point), the normalization condition for the trace of $\rho_A$ is $1=\sum\limits_{i=1}^{N_g}e^{-\xi_i}=N_ge^{-\xi}$, where $\xi_i$ are the entanglement eigenvalues. The EE is then $S(N_A)=\mathrm{ln}(N_g)$. In the large $N_p$ limit, we obtain a linear, {\it i.e.} volume scaling for the EE, different from the standard area law seen in real space systems \cite{RevModPhys.82.277}. (This volume scaling hints at a drawback of momentum space DMRG from the view of general information theory. Still, the structure of the ES might render this formulation advantageous in the end, for the range of finite systems which are available.) By varying $N_A$ and expanding the EE in the large $N_p$ limit around $N_A=\frac{N_P}{2}$, we find $S_A(N_p,N_A)=S_A(N_A=\frac{N_p}{2})-\frac{2}{N_p}(\frac{N_p}{2}-N_A)^2$. This behavior at $\Delta=1$ is shown in Fig.~\ref{Flat}.

Despite similarity in state counting to the Laughlin ES on the quantum Hall sphere, one important difference is that the ES at the Heisenberg point is flat, consistent with the Heisenberg point being non-chiral. (Note the ES of the FQH Laughlin state on the sphere mimics a linearly dispersing chiral $U(1)$ mode in the conformal limit \cite{Thomale_AC:prl10}.) An important consequence of the flat spectrum at the Heisenberg point is the absence of TEE, which we have numerically confirmed, whereas the FQH Laughlin state on the sphere has TEE \cite{PhysRevLett.98.060401}.

{\bf \em Bosonic Entanglement Gap}
-- We now vary $\Delta$ and investigate the behavior of the EG.  Adjusting $\Delta$ varies the scaling dimension of the underlying CFT in the gapless phase. As stated before, the EG is infinite at the HS point (which can be thought of as an $SU(2)$ invariant deformation away from the Heisenberg point), at which the fractionalized excitations, spinons, are free and interact only through their mutual statistics. Moving from the HS point to the Heisenberg point introduces interactions between spinons and dresses the state, but the EG still remains in the TD limit. The EG for bosons is defined as the minimal difference between the generic entanglement levels and the low lying universal levels. Our conclusions do not depend on whether we define the EG as a direct gap constrained to a given sector $(N_A, M_A)$ or as a global gap over all $M_A$. We find that the EG for bosons is not open throughout the entire gapless region. This is not affected by considering an approximate decomposition in total momentum or an exact decomposition in crystal momentum.

Fig.~\ref{Boson_ES_22} shows representative plots of the bosonic ES as a function of $\Delta$. The EG decreases as $\Delta$ is lowered from $1$. Qualitatively, this is due to interactions developing between spinons, which further dress the approximate product-like spinon state present at the Heisenberg point. The EG for 22 sites closes at $\Delta\approx\frac{1}{4}$. Finite size scaling suggests that in the TD limit the EG closes at $\Delta\approx\frac{3}{4}$ (Fig.~\ref{GAP}). The exact location of the closure is beyond the scope of present work. The important feature is the closure of the EG for some value of $\Delta$ in the gapless regime. This is seen {\it even} for finite system sizes. It might allow for the interpretation that the momentum space EG is not, by itself, a necessary signature for a gapless S=1/2 spin fluid state.

Note real space symmetries also yield fingerprints in the momentum space ES for bosons. For $\Delta=0$, one can unitarily transform the Hamiltonian from negative $J_1$ to positive $J_1$ by a $\pi$ rotation about the $z$ axis on every second site. We analytically show in the supplemental material this leads to a shift in momentum by half a lattice vector and a reflection in the ES. We also prove a reflection seen in the bosonic ES when $J_1=0,J_2\neq0$.

{\bf \em Fermionic ES} --
\begin{figure*}
\subfloat[][$\Delta=1$]{
 \includegraphics[width=.33\linewidth]{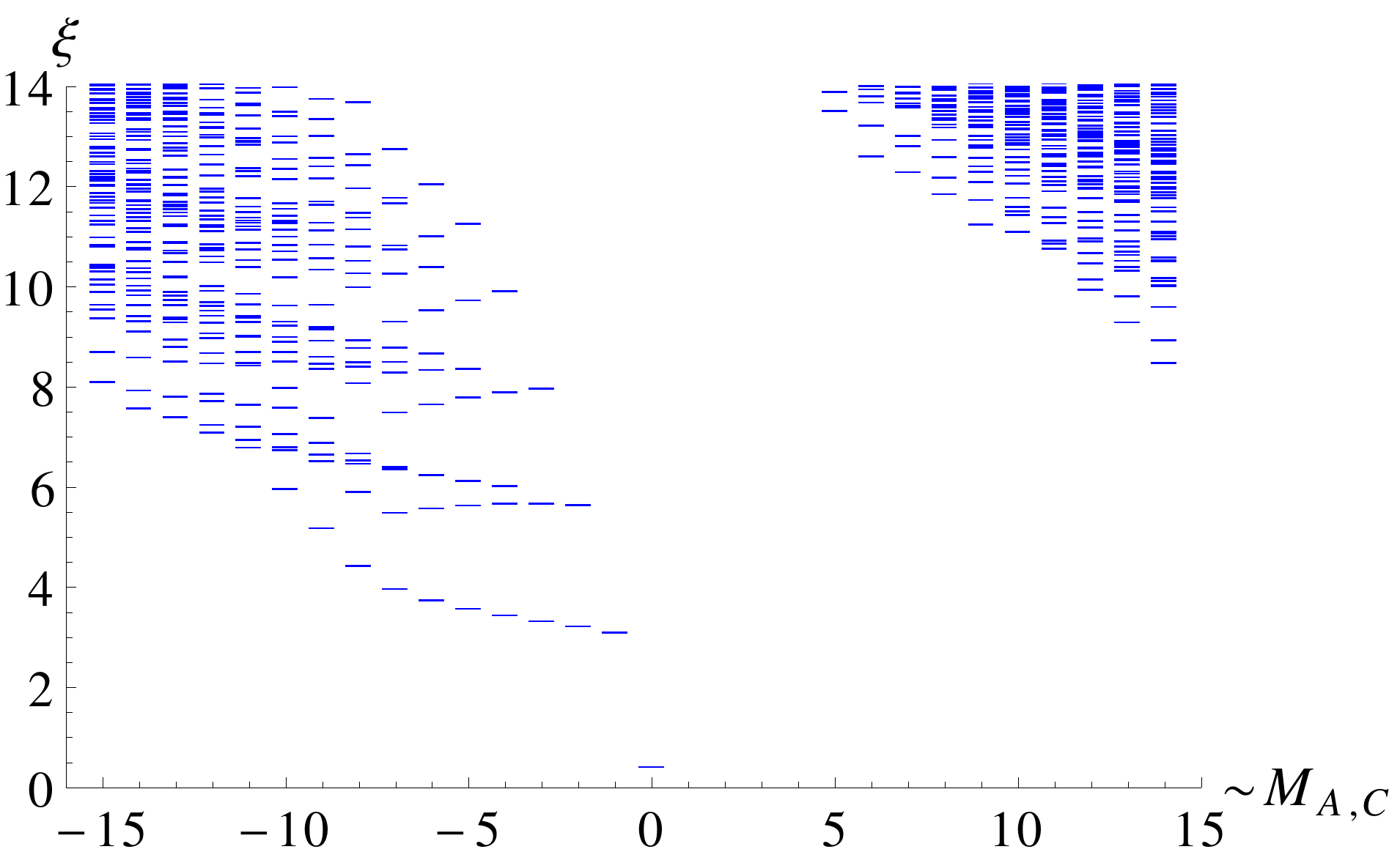}
 \label{Heisenberg}
}
\subfloat[][$\Delta=2$]{
 \includegraphics[width=.33\linewidth]{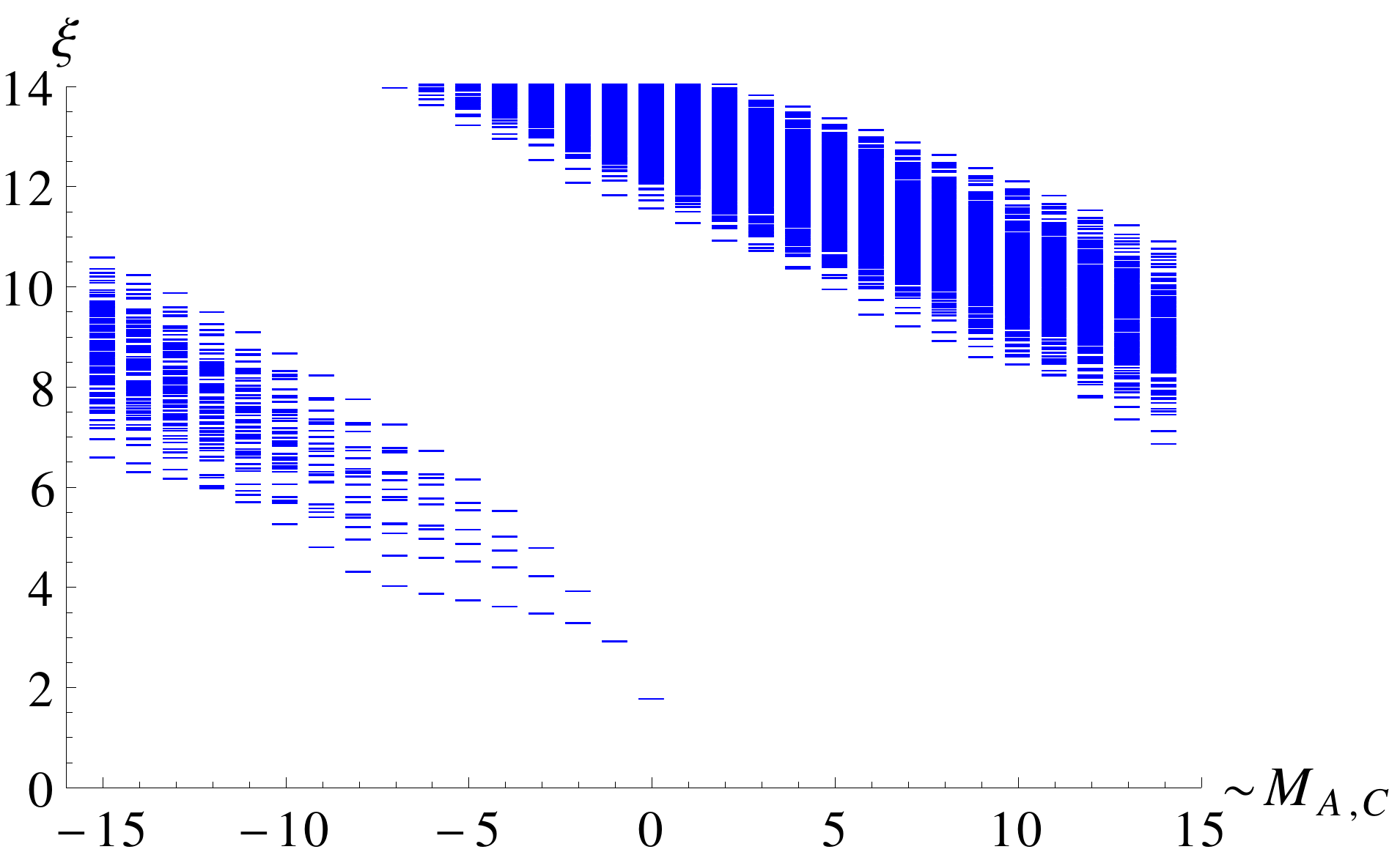}
 \label{fermion_heisenberg}
}
\subfloat[][$\Delta=7$]{
 \includegraphics[width=.33\linewidth]{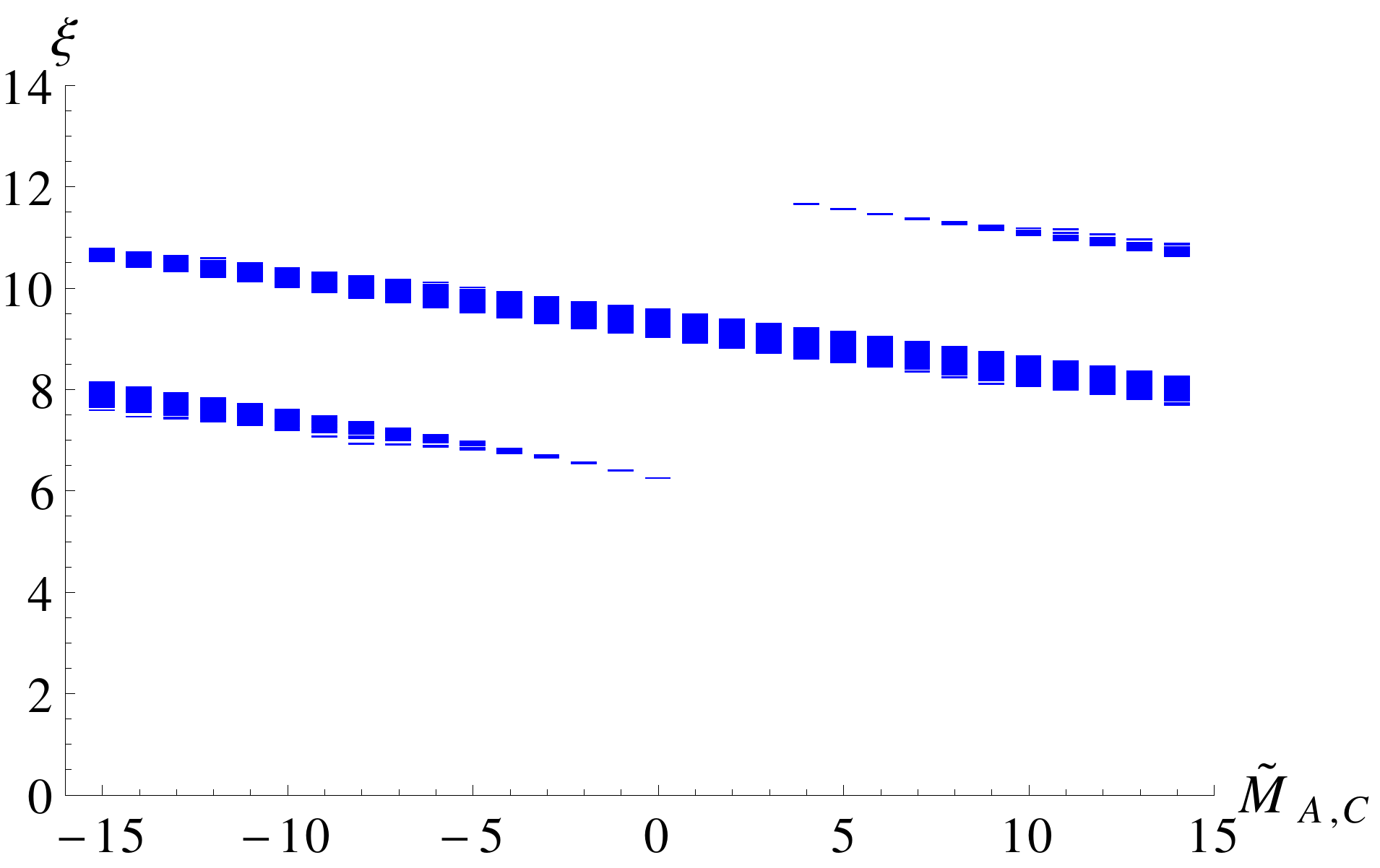}
 \label{fermions_seven}
}
\caption{(color online) Fermionic ES for representive values of $\Delta$, with $N=30$ and $N_A=7$. The fermionic ES remains qualitatively similar for $\Delta=1$ and $\Delta=2$ despite a phase transition at $\Delta=1$ and is linear for $\Delta=7$. The entanglement eigenvalues $\xi$ are plotted versus $\tilde{M}_{A,C}=M_{A,C}+13$. We have shifted the Brillouin zone by $M_{C}=13$ to have the lowest level at $\tilde{M}_{A,C}=0$.}
\label{Fermion_ES}
\end{figure*}
Fig.~\ref{Fermion_ES} shows plots of the fermionic ES for representative values of $\Delta$. We first observe that the counting of 1,1,2,3,5... for all values of $\Delta$ (Fig.~\ref{Fermion_ES}), starting from $\tilde{M}_{C,A}=0$, is trivial, as it links to total Hilbert space state counting in the respective momentum sector. This highlights the importance of the $U(1)$ counting seen for bosons near $\Delta=1$ where the Hilbert space is not exhausted by this counting.

At $\Delta=0$, the fermionic left and right movers are not entangled and form a product state \cite{PhysRevA.89.032326,2010AnPhy.325..924B} with an infinite EG. Qualitatively the gap decreases as we increase $\Delta$ from zero due to interactions developing between fermions, complementary to what is observed for bosons. The EG captures the low energy properties of $H_e$, thus phase transitions in $H_e$. Fig.~\ref{GAP} shows the TD projection of the fermionic EG. The EG remains relatively large and finite well past $\Delta=1$. The inset of Fig.~\ref{GAP} shows that the scaling with inverse system size always decays linearly and the gap remains open in the TD limit past $\Delta=1$. For $\Delta\gtrsim 1.7$, the system sizes we consider are larger than the correlation length \cite{PhysRevB.66.235108}, so we expect our scaling to be reliable. For example, the correlation length for the charge stiffness, taken from Ref.~\cite{PhysRevB.66.235108}, is $\approx1.5$ lattice spacings for $\Delta=3$, while the largest system size we consider is 10 times bigger (dividing $N$ by two due to PBCs).  This should also be compared to the scaling difference between the two lowest physical energy levels. Analyzing the scaling of the difference in the two lowest physical energy levels in the $XXZ$ model, one can accurately detect the phase transition at $\Delta=1$ with only ten sites \cite{PhysRev.135.A640}. As such, the system sizes considered here are large enough in principle to detect phase transitions near the TD value. We conclude that the EG and the ES systematically do not capture the phase transition from the gapless to gapped phase.

The ES deep in the Ising phase is linear as seen in Fig.~\ref{fermions_seven} \cite{Supplemental}. This adds to the argument that the EG extends past $\Delta=1$. The reasoning is as follows: Starting near the transition (on the gapped side) of $H_e$ (at zero entanglement temperature, $T_e$) we can imagine increasing $T_e$. Increasing $T_e$, we expect the gapped phase to become gapless due to thermal fluctuations, thus giving us the required physical phase diagram at $T_e=1$ \cite{2013arXiv1311.2946C}.

{\bf \em Conclusions} -- We have studied the momentum space ES for both bosons and fermions. We have shown with analytical methods and exact diagonalization the momentum space ES fails to detect physical phase transitions. More explicitly, the EG seen for bosons \cite{PhysRevLett.105.116805} {\it does not} remain open for arbitrary scaling dimensions in the $c=1$ CFT domain. For fermions, we found that the low energy (highly entangled) part of $H_e$ does not host a phase transition near a corresponding physical phase transition. Our findings for fermions suggest the results of Ref.~\cite{disorder_fermi_MS},  which state that momentum space ES can characterize disordered one-dimensional fermionic systems, need to be taken with caution in the presence of interactions. 

Our results are useful for numerical techniques such as DMRG, where one discards states of $\rho_A$ with low entanglement. For both fermions and bosons, a large separation of scales in entanglement persists in a relatively large region around the phase transition at $\Delta=1$. Assuming there is still enough entanglement weight located below the EG, this might, despite a volume law for EE, allow a momentum space based DMRG code to probe the critical point and the region around it. This work highlights that the non-universality of the ES pointed out in Ref.~\cite{2013arXiv1311.2946C} might in certain cases establish a {\em useful} feature for numerical applications. 

{\bf \em Acknowledgments} -- We thank V. Chua, S. Furukawa, M. Oshikawa, D. Lorshbough, and P. Laurell for useful discussions. RT particularly thanks D. Arovas and B. A. Bernevig for discussions and collaborations on related topics. RL was supported by NSF Graduate Research Fellowship award number 2012115499. RL thanks the hospitality of the Insitute for Theoritical Physics, Univesity of W\"{u}rzburg and University of Tokyo where part of this work was completed under NSF EAPSI award number OISE-1309560 and Japan Society for the Promotion of Science (JSPS) Summer Program 2013. AML acknowledges support through FOR1807 (DFG / FWF). GAF acknowledges financial support through ARO Grant No. W911NF-09-1-0527 and NSF Grant No. DMR-0955778. MG and RT are supported by the ERC starting grant TOPOLECTRICS of the European Research Council (ERC-StG-Thomale-2013-336012). The authors acknowledge the Texas Advanced Computing Center (TACC) at The University of Texas at Austin for providing computing resources that have contributed to the research results reported within this paper. URL: http://www.tacc.utexas.edu
\appendix
\section{Supplementary Material I: Detailed Description of Transformed Hamiltonians}
For completeness, we provide a detailed description of the $XXZ$ Hamiltonian reformulated via bosons and fermions in momentum space. The transformed bosonic nearest neighbor Hamiltonian is given by
\begin{align}
H = \sum_{i=1}^N \Biggl(J(b^{\dagger}_ib_{i+1}^{\phantom{\dagger}}+b^{\dagger}_{i+1}b_{i}^{\phantom{\dagger}}+\Delta(n_i-\frac{1}{2})(n_{i+1}-\frac{1}{2}))+\nonumber \\
V(n_{i}-\frac{1}{2})(n_{i}-\frac{1}{2})\Biggr),
\end{align}
where $V$ is the hardcore potential and $n_i=b^{\dagger}_ib_i^{\phantom{\dagger}}$. The mapping between hardcore bosons and spins is exact as $V\rightarrow \infty$. After inserting the Fourier transform of the bosonic operators defined in the main text, the Hamiltonian becomes
\begin{align}
&H=\sum_k (J \cos(k)-V- J \Delta) b^{\dagger}_kb_k^{\phantom{\dagger}}+\nonumber\\
&\frac{1}{N}\sum_{\substack{k_1,k_2,\\k_3,k_4}}\Biggl( \Bigl(V+
J \Delta \cos(k_3-k_4)\Bigr)b^{\dagger}_{k_1}b_{k_2}^{\phantom{\dagger}}b^{\dagger}_{k_3}b_{k_4}^{\phantom{\dagger}}\delta_{-k_1+k_2-k_3+k_4,0}\Biggr).
\end{align}

The transformed nearest neighbor fermionic Hamiltonian is given by 
\begin{align}
H = J \sum_{i=1}^N \left(c^{\dagger}_ic_{i+1}^{\phantom{\dagger}}+c^{\dagger}_{i+1}c_{i}^{\phantom{\dagger}}+\Delta(n_i-\frac{1}{2})(n_{i+1}-\frac{1}{2})\right).
\end{align}
After inserting the Fourier transform of the fermionic operators defined in the main text, the Hamiltonian becomes
\begin{align}
&H=J \, \Biggl( \, \sum_k (\cos(k)-\Delta) c^{\dagger}_kc_k^{\phantom{\dagger}}+\nonumber \\
&\frac{\Delta}{N}\sum_{\substack{k_1,k_2,\\k_3,k_4}}\Bigl(\cos(k_3-k_4)c^{\dagger}_{k_1}c_{k_2}^{\phantom{\dagger}}c^{\dagger}_{k_3}c_{k_4}^{\phantom{\dagger}}\delta_{-k_1+k_2-k_3+k_4,0}\Bigr) \, \Biggr). 
\end{align}

\section{Supplementary Material II: Entanglement Spectrum of The Ising Phase for Fermions}
\begin{figure}
\begin{center}
 \includegraphics[width=1\linewidth]{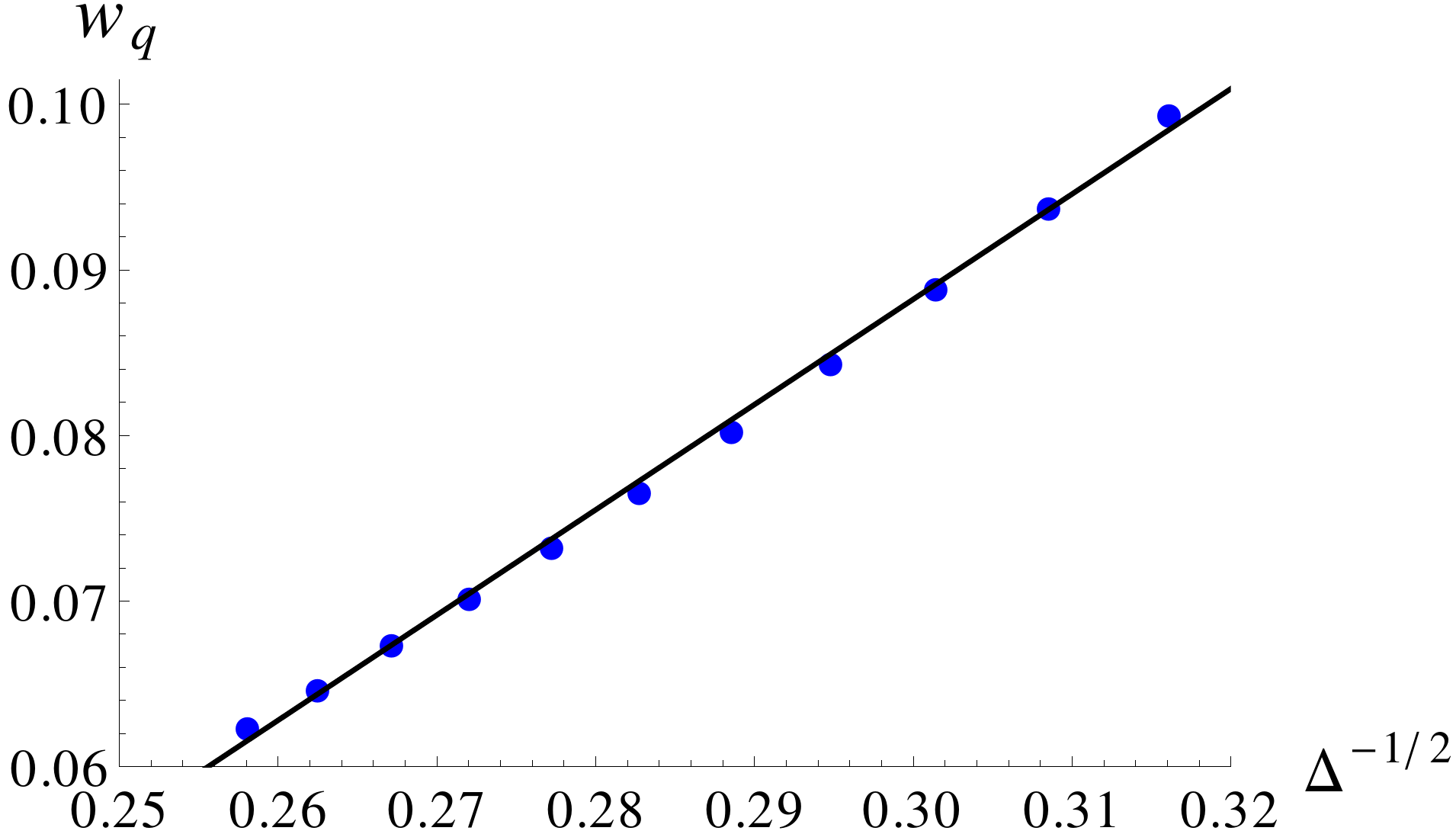}
\end{center}
\caption{(color online)  The slope of the fermionic ES versus $\frac{1}{\sqrt{\Delta}}$. All data taken from 30 sites, with a cut region containing 7 fermions. The blue dots represent the numerical value of the ES slope and the black line is a linear fit.}
\label{Slope}
\end{figure}

Here we provide a proof that the ES of the Ising phase for fermions is linear. The bosonized version of the $XXZ$ Hamiltonian \cite{Gogolin:book} is given by 
\begin{equation}
H=\frac{v}{2}\int \mathrm{d}x\Biggl( \Pi^2+(1+\frac{4\Delta}{\pi})(\partial_x \phi)^2+ \frac{2\Delta}{(\pi a)^2} \cos (\sqrt{16\pi}\phi)\Biggr),
\end{equation}
where $a$ is the short distance cutoff and $v$ is the velocity. We will transform to fermions later. We now expand the cosine in terms of the scalar fields to second order, valid for large $\Delta$ (deep in Ising phase) where $\phi$ has small fluctuations about a value that minimizes the cosine term.  After dropping an overall constant, the Hamiltonian is given by
\begin{equation}
H=\frac{v}{2} \int \mathrm{d}x\Biggl(  (\Pi)^2+(1+\frac{4\Delta}{\pi})(\partial_x \phi)^2+\frac{16\pi\Delta}{(\pi a)^2}\phi^2 \Biggr).
\end{equation}
Plugging in the following mode expansions (ignoring zero modes) for $\phi$ and $\Pi$
\begin{subequations}
\begin{gather}
\phi=\sum_{q\neq0} \frac{1}{\sqrt{2|q|N}}e^{-iqx}(a_q^{\dagger}+a_{-q}^{\phantom{\dagger}}),\\
\Pi=\sum_{q\neq0} \sqrt{\frac{|q|}{2N}}e^{-iqx}(a_q^{\dagger}-a_{-q}^{\phantom{\dagger}}),
\end{gather}
\end{subequations}
the Hamiltonian takes the form
\begin{align}
 H=&\frac{v}{2}\sum_{q\neq0} 
  \left( a_q^{\dagger},a_{-q}^{\phantom{\dagger}} \right) 
  \begin{pmatrix} A_q & B_q\\ B_q & A_q \end{pmatrix}  
  \begin{pmatrix} a_q^{\phantom{\dagger}} \\ a_{-q}^{\dagger} \end{pmatrix} ,
\end{align}
where
\begin{equation}
A_q=(1+\frac{2\Delta}{\pi})|q|+\frac{8\Delta}{\pi a^2|q|},
\end{equation}
and
\begin{equation}
B_q=\frac{2\Delta}{\pi}(|q|+\frac{4}{a^2|q|}).
\end{equation}
We can diagonalize $H$ via a Bogoliubov transformation 
\begin{equation}
 \begin{pmatrix} a_q^{\phantom{\dagger}} \\ a_{-q}^{\dagger} \end{pmatrix}
 =\begin{pmatrix} \cosh \theta_q &  \sinh \theta_q \\ \sinh \theta_q & \cosh \theta_q\\ \end{pmatrix}  
 \begin{pmatrix} b_q^{\phantom{\dagger}} \\ b_{-q}^{\dagger} \end{pmatrix},
\end{equation}
with
\begin{subequations}
\begin{gather}
 \cosh \left( 2\theta_q \right) = \frac{A_q}{\lambda_q},~~\sinh \left( 2\theta_q \right)=-\frac{B_q}{\lambda_q},\\
 \lambda_q^2=A_q^2-B_q^2=(1+\frac{4\Delta}{\pi})|p|^2+\frac{16\Delta}{\pi}.
\end{gather}
\end{subequations} 
This yields the diagonal bilinear form
\begin{equation}
H=v\sum_{q\neq0}\lambda_q(b_{q}^{\dagger}b_q^{\phantom{\dagger}}+\frac{1}{2}).
\end{equation}
The ground state $|0\rangle$ of $H$ is specified by the condition that $b_q|0\rangle=0$ $\forall$ $q$. From this, we can calculate the entanglement Hamiltonian using free field theory methods \cite{2003JPhA...36L.205P}. We first calculate the two-point correlation functions for the right movers with right-moving momentum $q>0$. Using the ground state $|0\rangle$ of $H$, 
\begin{align}\label{eq:akdak}
 \langle0|a_q^{\dagger}a_q^{\phantom{\dagger}}|0\rangle &= \sinh^2 \theta_q = \frac{\cosh(2\theta_q)-1}2. 
\end{align}
We introduce the ansatz
\begin{equation}
 \rho_A = \frac{e^{-H_e}}{Z_e},\;
 Z_e= \mathrm{Tr} e^{-H_{e}},
\end{equation}
with 
\begin{equation}\label{eq:He_osc_ch}
 H_e= \sum_{q>0} w_q \left(a_q^{\dagger}a_q^{\phantom{\dagger}} +\frac12 \right). 
\end{equation}
This gives the Bose distribution 
\begin{equation}\label{eq:akdak_ansatz}
 \mathrm{Tr}\left( a_q^\dagger a_q^{\phantom{\dagger}} \rho_A \right) = \frac1{e^{w_q}-1}. 
\end{equation}
Equating Eq.~\eqref{eq:akdak_ansatz} with Eq.~\eqref{eq:akdak}, we obtain the expression of $w_q$ as
\begin{equation}
 w_q=\ln \frac{\cosh(2\theta_q)+1}{\cosh(2\theta_q)-1}.
\end{equation}
One must divide this slope by $\pi$ to obtain the fermionic ES slope \cite{Giamarchi:book}. In the small $q$ limit, after setting $a=1$, we obtain
\begin{equation}
w_q=\frac{1}{\sqrt{\pi}}\frac{1}{\sqrt{\Delta}}|q|,
\label{es_slope}
\end{equation}
which indeed is a linear spectrum. Numerical evidence for the behaviour of the ES slope as a function of $\Delta$ is shown in Fig.~\ref{Slope}. The numerical slope obtained is of the form $c+\frac{b}{\sqrt{\Delta}}$, where $c\sim-.10$ and $b\sim.63$, which is in quantitative agreement with the theoretical prediction given by Eq.~\eqref{es_slope}.

\section{Supplementary Material III: Symmetries for Bosons}


\begin{figure*}
\subfloat[][$J_1=-4,J_2=1,\Delta=0$]{
 \includegraphics[width=.33\linewidth]{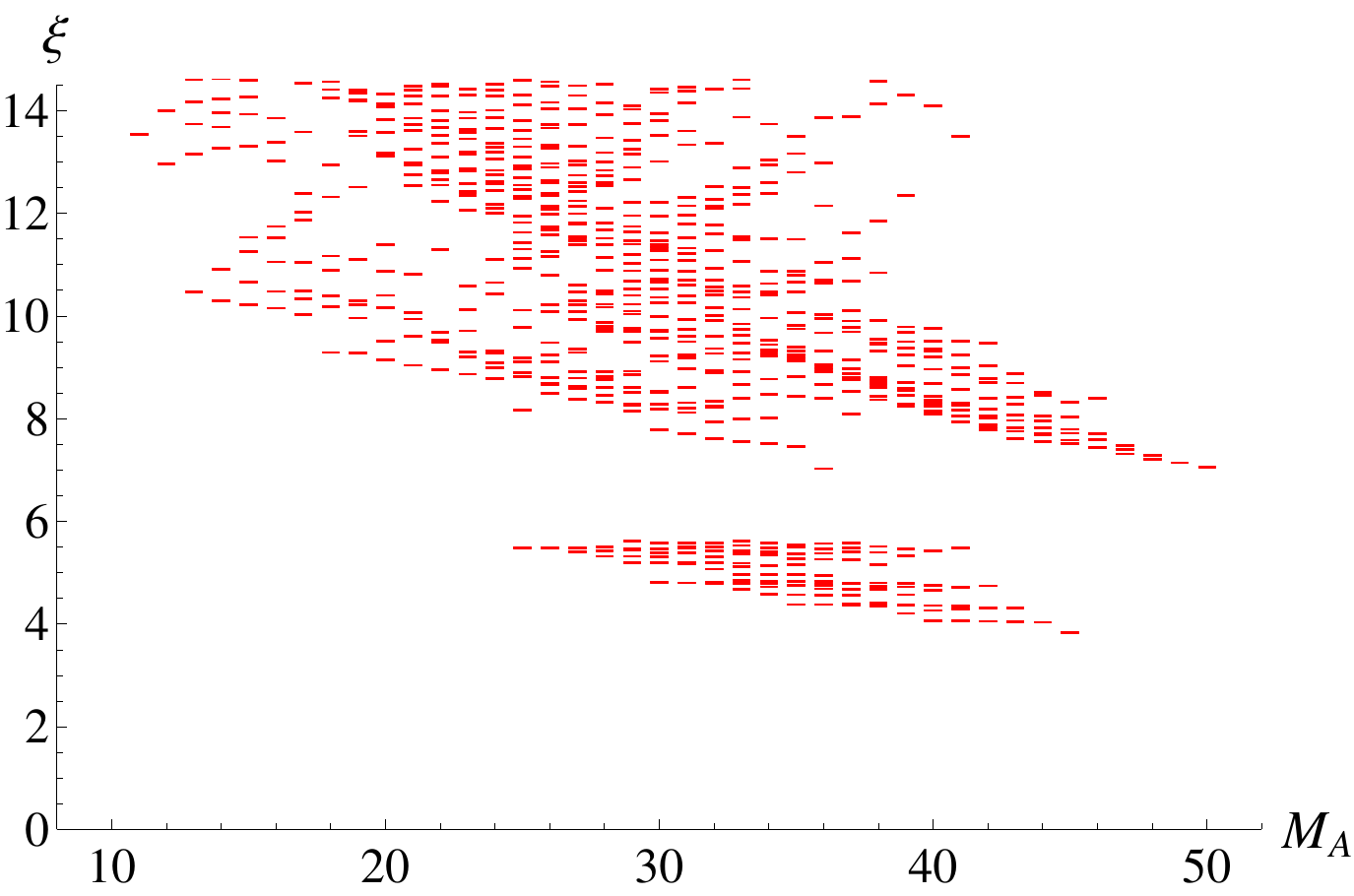}
 \label{$J_1negative$}
}
\subfloat[][$J_1=4,J_2=1,\Delta=0$]{
 \includegraphics[width=.33\linewidth]{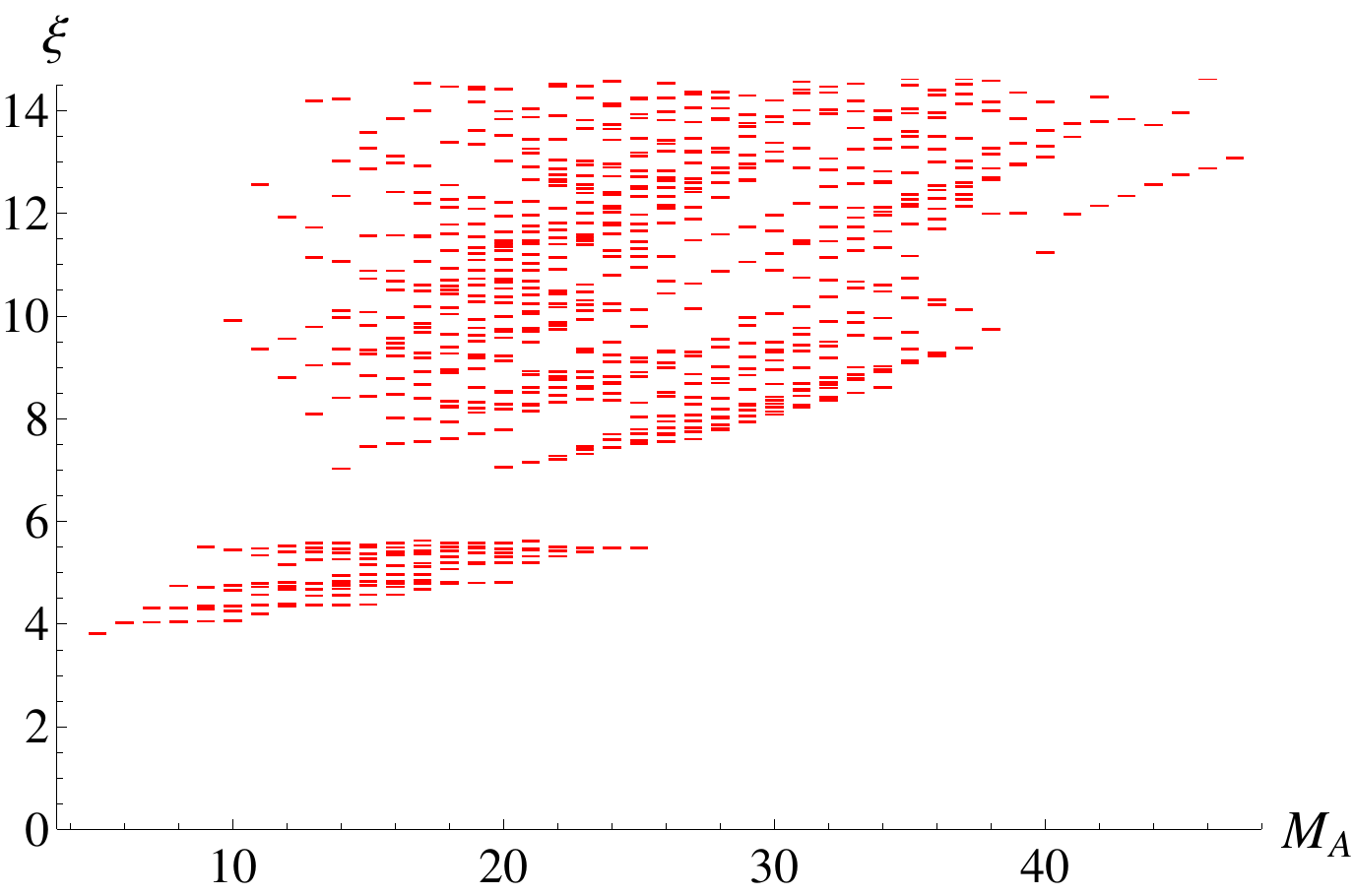}
 \label{$J_1positive$}
}
\subfloat[][$J_1=0,J_2=1,\Delta=0.2$]{
 \includegraphics[width=.33\linewidth]{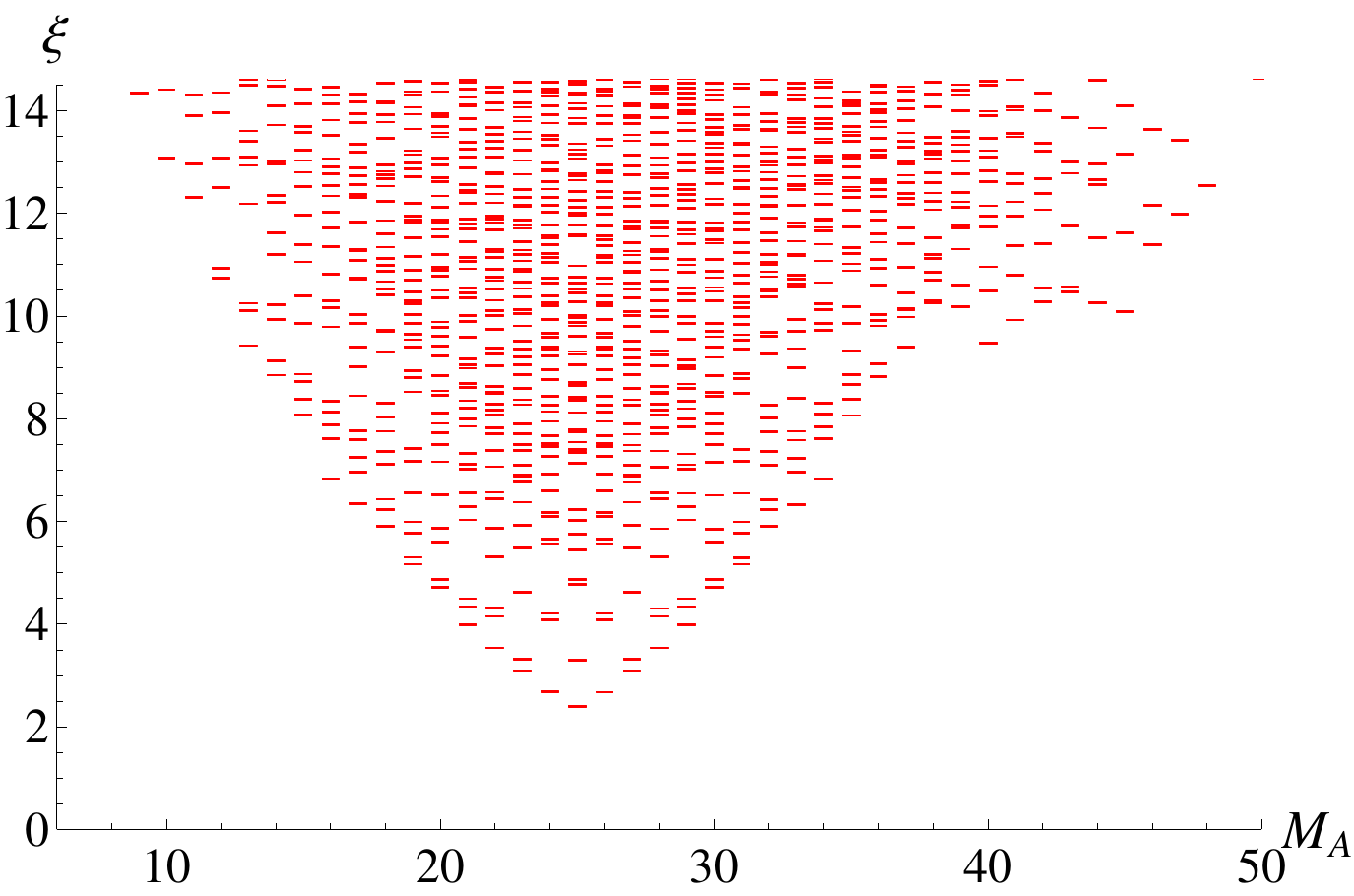}
 \label{J_1=0}
}
\caption{(color online) Bosonic ES for system size $N=20$ for various values of $J_1,J_2$ and $\Delta$, with a cut region containing 5 magnons. The reflection properties observed are analytically proven in Supplementary Material III: Symmetries for Bosons.}
\label{Boson_ES_20}
\end{figure*}

In this section, we prove the effect of certain real-space symmetries on the momentum space ES.
\newline
\newline
{\bf \em Case 1: $\Delta=0$} -- For $\Delta=0$, one can unitarily transform the Hamiltonian from negative $J_1$ to positive $J_1$ by a $\pi$ rotation about the $z$ axis on every second site. This provides a relationship between the real-space wave-functions for positive and negative $J_1$. This relationship is given by
\begin{equation}
\psi^-(z_{j_1}...z_{j_{\kappa}})=\psi^+(z_{j_1}...z_{j_{\kappa}})e^{i\pi (j_1 +1)}...e^{i\pi( j_{\kappa}+1)}.
\label{eq:shift_negative_coeff}
\end{equation}
where $\psi^{+}(z_{j_1}...z_{j_{\kappa}})$ is the ground state coefficient of a given spin configuration for the case when $J_1$ is positive, $\psi^{-}(z_{j_1}...z_{j_{\kappa}})$ is the ground state coefficient of a given spin configuration when $J_1$ is negative, $z_{j_i}$ is the location of the $i$th down spin and $\kappa=\frac{N}{2}$ is the total number of down spins. We first see how this affects the momentum space wave-function. In terms of the real space coefficients, the momentum space coefficients \cite{PhysRevLett.105.116805} for negative $J_1$ are given by
\begin{align}
\psi^-(m_1...m_{\kappa})=&\sum\limits_{j_1...j_{\kappa}} e^{i\frac{2\pi}{N}j_1m_1}...e^{i\frac{2\pi}{N}j_{\kappa}m_{\kappa}}\psi^-(z_{j_1}...z_{j_{\kappa}})=\nonumber \\
&\sum\limits_{j_1...j_{\kappa}} e^{i\frac{2\pi}{N}j_1(m_1+\frac{N}{2})}...\psi^+(z_{j_1}...z_{j_{\kappa}}).
\label{eq:shift_negative}
\end{align}
We see that we now have an effective momentum which appears in the Fourier transform. The effective momentum for each spin-flip is $m_{eff}=m+\frac{N}{2}$.

To see how this affects the ES, we first write the ground state wave function in a Schmidt decomposition of a fixed number of left and right moving particles as
\begin{equation}
\left|\psi^{-}\right\rangle = \sum_{i=0}^N \sum_{\alpha,\beta} \sum_{p+q=iN} \psi^{-}_{i; \alpha p; \beta q}
\left|\alpha p \right\rangle \left|\beta q \right\rangle,
\end{equation}
where $\alpha$ and $\beta$ represent all possible occupation states that give rise to a subsystem with left movers with total momentum $p$ and right movers with total momentum $q$, and the sum over $i$ represents the sum over all possible momentum sectors. Using the approximation that $\left\langle \beta q \middle| \beta' q' \right\rangle \sim \delta_{pp'} \delta_{\beta\beta'}$ we arrive at the reduced density matrix for negative $J_1$
\begin{equation}
\rho^{-}_A = \sum_{i,i',p,\alpha,\alpha',\beta} \psi^{*;-}_{\substack{i';\alpha'p;\\\beta (i'N-p)}}\psi^{-}_{\substack{i;\alpha p;\\\beta (iN-p)}}
\left|\alpha\phantom{'} \! p \right\rangle \left\langle \alpha'p \right|.
\label{eq:red_density}
\end{equation}
Substituting Eq. \eqref{eq:shift_negative} into Eq. \eqref{eq:red_density} produces the reduced density matrix for negative $J_1$ in terms of the ground-state coefficients for positive $J_1$,
\begin{align}
\label{eq:rho_Ising}
\rho^{-}_A = \sum_{p,\alpha,\alpha',\beta}\psi^{*;+}_{\substack{\alpha'p+\frac{N N_A}{2};\\\beta (\kappa^2-p-\frac{N N_A}{2})}}\psi^{+}_{\substack{\alpha p+\frac{N N_A}{2};\\\beta (\kappa^2-p-\frac{N N_A}{2})}}
\left| \alpha\phantom{'} \! p \right\rangle \left\langle \alpha' p \right|,
\end{align}
where $N_A$ is the number of particles in region $A$, which is equal to the number of particles in region $B$.
Here we have used the fact that most of the weight is in the $\kappa^2$ sector for the positive case which allows us to collapse the sum over momentum sectors. Thus, we see that the entanglement levels for negative $J_1$ are the same as positive $J_1$, but appear at momenta shifted by $\frac{NN_A}{2}$. 
This completes the proof for the reflection observed in Figures~\ref{$J_1negative$} and~\ref{$J_1positive$}.
\newline
{\bf \em Case 2: $J_1=0$, $J_2\neq0$} -- We close this section by explaining the reflection about $M_A=25$ as seen in the bosonic ES in Fig.~\ref{J_1=0}, around $J_1=0$ and positive $J_2$. For $J_1=0$ we are in a dimerized phase. As a consequence, for an even number of sites, one can split the system into subsystems, i.e. even and odd sites. In addition to the total spin of the system being zero, the total spin of each sub-system is zero for the ground state. This fact is crucial to this proof. One can write the real-space wave-function as
\begin{align}
&\left| \psi \right\rangle=\sum_{j_1,\:\! \ldots, j_{\frac{\kappa}{2}}}^{\frac{N}{2}} \sum_{j'_1,\:\! \ldots , j'_{\frac{\kappa}{2}}=1}^{\frac{N}{2}} \psi(2j_1,\:\! \ldots , 2j_{\frac{\kappa}{2}};2j'_1-1,\:\! \ldots , 2j'_{\frac{\kappa}{2}}-1)\nonumber \\
&S_{2j_1}^-\:\! \ldots S_{2j_{\frac{\kappa}{2}}}^-S_{2j'_1-1}^-\:\! \ldots S_{2j'_{{\frac{\kappa}{2}}}-1}^-\left| F \right\rangle,
\end{align}
where the primes indicate the sum over the odd sites and the unprimed $j$'s are on even sites. The coefficients of the Fourier transform are given by

\begin{align}
&\psi(m_1  \:\! \ldots \,  m_k) = \sum_{\substack{j_1,\:\! \ldots , j_{\frac{\kappa}{2}}=1\\j'_1,\:\! \ldots ,  j'_{\frac{\kappa}{2}}=1}}^{\frac{N}{2}} \psi(2j_1,\:\! \ldots , 2j_{\frac{\kappa}{2}};2j'_1-1,\:\! \ldots , 2j'_{\frac{\kappa}{2}}-1)\nonumber \\
&e^{\frac{i2\pi}{N}m_12j_1} \:\! \ldots \, e^{\frac{i2\pi}{N}m_12j_\kappa} e^{\frac{i2\pi}{N}m_1(2j'_1-1)}\:\! \ldots \,e^{\frac{i2\pi}{N}m_1(2j'_{\frac{\kappa}{2}}-1)}
\end{align}

Taking $m$ to $m+N/2$ on even sites simply returns a factor of $e^{i2\pi}$. Doing this for the odd sites returns a factor of $e^{-i\pi}$. However, because we have a fixed number of down spins on the odd sites, this factor is the same for every term in the sum and just returns an overall phase. In this case it is unity, since there are a even number of down spins on the odd sites. Thus the momentum space wave-function is invariant under $m \mapsto m+N/2$. We have seen earlier this leads to a reflection in the entanglement spectrum about a certain $M_A$ and thus explains the observed reflection in the entanglement spectrum for $J_1=0$ (Fig.~\ref{J_1=0}).


\begin{thebibliography}{85}%
\makeatletter
\providecommand \@ifxundefined [1]{%
 \@ifx{#1\undefined}
}%
\providecommand \@ifnum [1]{%
 \ifnum #1\expandafter \@firstoftwo
 \else \expandafter \@secondoftwo
 \fi
}%
\providecommand \@ifx [1]{%
 \ifx #1\expandafter \@firstoftwo
 \else \expandafter \@secondoftwo
 \fi
}%
\providecommand \natexlab [1]{#1}%
\providecommand \enquote  [1]{``#1''}%
\providecommand \bibnamefont  [1]{#1}%
\providecommand \bibfnamefont [1]{#1}%
\providecommand \citenamefont [1]{#1}%
\providecommand \href@noop [0]{\@secondoftwo}%
\providecommand \href [0]{\begingroup \@sanitize@url \@href}%
\providecommand \@href[1]{\@@startlink{#1}\@@href}%
\providecommand \@@href[1]{\endgroup#1\@@endlink}%
\providecommand \@sanitize@url [0]{\catcode `\\12\catcode `\$12\catcode
  `\&12\catcode `\#12\catcode `\^12\catcode `\_12\catcode `\%12\relax}%
\providecommand \@@startlink[1]{}%
\providecommand \@@endlink[0]{}%
\providecommand \url  [0]{\begingroup\@sanitize@url \@url }%
\providecommand \@url [1]{\endgroup\@href {#1}{\urlprefix }}%
\providecommand \urlprefix  [0]{URL }%
\providecommand \Eprint [0]{\href }%
\providecommand \doibase [0]{http://dx.doi.org/}%
\providecommand \selectlanguage [0]{\@gobble}%
\providecommand \bibinfo  [0]{\@secondoftwo}%
\providecommand \bibfield  [0]{\@secondoftwo}%
\providecommand \translation [1]{[#1]}%
\providecommand \BibitemOpen [0]{}%
\providecommand \bibitemStop [0]{}%
\providecommand \bibitemNoStop [0]{.\EOS\space}%
\providecommand \EOS [0]{\spacefactor3000\relax}%
\providecommand \BibitemShut  [1]{\csname bibitem#1\endcsname}%
\let\auto@bib@innerbib\@empty
\bibitem [{\citenamefont {Amico}\ \emph {et~al.}(2008)\citenamefont {Amico},
  \citenamefont {Fazio}, \citenamefont {Osterloh},\ and\ \citenamefont
  {Vedral}}]{RevModPhys.80.517}%
  \BibitemOpen
  \bibfield  {author} {\bibinfo {author} {\bibfnamefont {L.}~\bibnamefont
  {Amico}}, \bibinfo {author} {\bibfnamefont {R.}~\bibnamefont {Fazio}},
  \bibinfo {author} {\bibfnamefont {A.}~\bibnamefont {Osterloh}}, \ and\
  \bibinfo {author} {\bibfnamefont {V.}~\bibnamefont {Vedral}},\ }\href
  {\doibase 10.1103/RevModPhys.80.517} {\bibfield  {journal} {\bibinfo
  {journal} {Rev. Mod. Phys.}\ }\textbf {\bibinfo {volume} {80}},\ \bibinfo
  {pages} {517} (\bibinfo {year} {2008})}\BibitemShut {NoStop}%
\bibitem [{\citenamefont {Li}\ and\ \citenamefont {Haldane}(2008)}]{Li:prl08}%
  \BibitemOpen
  \bibfield  {author} {\bibinfo {author} {\bibfnamefont {H.}~\bibnamefont
  {Li}}\ and\ \bibinfo {author} {\bibfnamefont {F.~D.~M.}\ \bibnamefont
  {Haldane}},\ }\href {\doibase 10.1103/PhysRevLett.101.010504} {\bibfield
  {journal} {\bibinfo  {journal} {Phys. Rev. Lett.}\ }\textbf {\bibinfo
  {volume} {101}},\ \bibinfo {pages} {010504} (\bibinfo {year}
  {2008})}\BibitemShut {NoStop}%
\bibitem [{\citenamefont {Thomale}\ \emph
  {et~al.}(2010{\natexlab{a}})\citenamefont {Thomale}, \citenamefont
  {Sterdyniak}, \citenamefont {Regnault},\ and\ \citenamefont
  {Bernevig}}]{Thomale_AC:prl10}%
  \BibitemOpen
  \bibfield  {author} {\bibinfo {author} {\bibfnamefont {R.}~\bibnamefont
  {Thomale}}, \bibinfo {author} {\bibfnamefont {A.}~\bibnamefont {Sterdyniak}},
  \bibinfo {author} {\bibfnamefont {N.}~\bibnamefont {Regnault}}, \ and\
  \bibinfo {author} {\bibfnamefont {B.~A.}\ \bibnamefont {Bernevig}},\ }\href
  {\doibase 10.1103/PhysRevLett.104.180502} {\bibfield  {journal} {\bibinfo
  {journal} {Phys. Rev. Lett.}\ }\textbf {\bibinfo {volume} {104}},\ \bibinfo
  {pages} {180502} (\bibinfo {year} {2010}{\natexlab{a}})}\BibitemShut
  {NoStop}%
\bibitem [{\citenamefont {Thomale}\ \emph {et~al.}(2011)\citenamefont
  {Thomale}, \citenamefont {Estienne}, \citenamefont {Regnault},\ and\
  \citenamefont {Bernevig}}]{PhysRevB.84.045127}%
  \BibitemOpen
  \bibfield  {author} {\bibinfo {author} {\bibfnamefont {R.}~\bibnamefont
  {Thomale}}, \bibinfo {author} {\bibfnamefont {B.}~\bibnamefont {Estienne}},
  \bibinfo {author} {\bibfnamefont {N.}~\bibnamefont {Regnault}}, \ and\
  \bibinfo {author} {\bibfnamefont {B.~A.}\ \bibnamefont {Bernevig}},\ }\href
  {\doibase 10.1103/PhysRevB.84.045127} {\bibfield  {journal} {\bibinfo
  {journal} {Phys. Rev. B}\ }\textbf {\bibinfo {volume} {84}},\ \bibinfo
  {pages} {045127} (\bibinfo {year} {2011})}\BibitemShut {NoStop}%
\bibitem [{\citenamefont {L\"auchli}\ \emph {et~al.}(2010)\citenamefont
  {L\"auchli}, \citenamefont {Bergholtz}, \citenamefont {Suorsa},\ and\
  \citenamefont {Haque}}]{Lauchli:prl10}%
  \BibitemOpen
  \bibfield  {author} {\bibinfo {author} {\bibfnamefont {A.~M.}\ \bibnamefont
  {L\"auchli}}, \bibinfo {author} {\bibfnamefont {E.~J.}\ \bibnamefont
  {Bergholtz}}, \bibinfo {author} {\bibfnamefont {J.}~\bibnamefont {Suorsa}}, \
  and\ \bibinfo {author} {\bibfnamefont {M.}~\bibnamefont {Haque}},\
  }\href@noop {} {\bibfield  {journal} {\bibinfo  {journal} {Phys. Rev. Lett.}\
  }\textbf {\bibinfo {volume} {104}},\ \bibinfo {pages} {156404} (\bibinfo
  {year} {2010})}\BibitemShut {NoStop}%
\bibitem [{\citenamefont {Rodriguez}\ \emph {et~al.}(2013)\citenamefont
  {Rodriguez}, \citenamefont {Davenport}, \citenamefont {Simon},\ and\
  \citenamefont {Slingerland}}]{PhysRevB.88.155307}%
  \BibitemOpen
  \bibfield  {author} {\bibinfo {author} {\bibfnamefont {I.~D.}\ \bibnamefont
  {Rodriguez}}, \bibinfo {author} {\bibfnamefont {S.~C.}\ \bibnamefont
  {Davenport}}, \bibinfo {author} {\bibfnamefont {S.~H.}\ \bibnamefont
  {Simon}}, \ and\ \bibinfo {author} {\bibfnamefont {J.~K.}\ \bibnamefont
  {Slingerland}},\ }\href {\doibase 10.1103/PhysRevB.88.155307} {\bibfield
  {journal} {\bibinfo  {journal} {Phys. Rev. B}\ }\textbf {\bibinfo {volume}
  {88}},\ \bibinfo {pages} {155307} (\bibinfo {year} {2013})}\BibitemShut
  {NoStop}%
\bibitem [{\citenamefont {Rodriguez}\ \emph {et~al.}(2012)\citenamefont
  {Rodriguez}, \citenamefont {Simon},\ and\ \citenamefont
  {Slingerland}}]{PhysRevLett.108.256806}%
  \BibitemOpen
  \bibfield  {author} {\bibinfo {author} {\bibfnamefont {I.~D.}\ \bibnamefont
  {Rodriguez}}, \bibinfo {author} {\bibfnamefont {S.~H.}\ \bibnamefont
  {Simon}}, \ and\ \bibinfo {author} {\bibfnamefont {J.~K.}\ \bibnamefont
  {Slingerland}},\ }\href {\doibase 10.1103/PhysRevLett.108.256806} {\bibfield
  {journal} {\bibinfo  {journal} {Phys. Rev. Lett.}\ }\textbf {\bibinfo
  {volume} {108}},\ \bibinfo {pages} {256806} (\bibinfo {year}
  {2012})}\BibitemShut {NoStop}%
\bibitem [{\citenamefont {Sterdyniak}\ \emph {et~al.}(2012)\citenamefont
  {Sterdyniak}, \citenamefont {Regnault},\ and\ \citenamefont
  {M\"oller}}]{PhysRevB.86.165314}%
  \BibitemOpen
  \bibfield  {author} {\bibinfo {author} {\bibfnamefont {A.}~\bibnamefont
  {Sterdyniak}}, \bibinfo {author} {\bibfnamefont {N.}~\bibnamefont
  {Regnault}}, \ and\ \bibinfo {author} {\bibfnamefont {G.}~\bibnamefont
  {M\"oller}},\ }\href {\doibase 10.1103/PhysRevB.86.165314} {\bibfield
  {journal} {\bibinfo  {journal} {Phys. Rev. B}\ }\textbf {\bibinfo {volume}
  {86}},\ \bibinfo {pages} {165314} (\bibinfo {year} {2012})}\BibitemShut
  {NoStop}%
\bibitem [{\citenamefont {Dubail}\ \emph
  {et~al.}(2012{\natexlab{a}})\citenamefont {Dubail}, \citenamefont {Read},\
  and\ \citenamefont {Rezayi}}]{Dubail:prb12}%
  \BibitemOpen
  \bibfield  {author} {\bibinfo {author} {\bibfnamefont {J.}~\bibnamefont
  {Dubail}}, \bibinfo {author} {\bibfnamefont {N.}~\bibnamefont {Read}}, \ and\
  \bibinfo {author} {\bibfnamefont {E.~H.}\ \bibnamefont {Rezayi}},\ }\href
  {\doibase 10.1103/PhysRevB.85.115321} {\bibfield  {journal} {\bibinfo
  {journal} {Phys. Rev. B}\ }\textbf {\bibinfo {volume} {85}},\ \bibinfo
  {pages} {115321} (\bibinfo {year} {2012}{\natexlab{a}})}\BibitemShut
  {NoStop}%
\bibitem [{\citenamefont {Jackson}\ \emph {et~al.}(2013)\citenamefont
  {Jackson}, \citenamefont {Read},\ and\ \citenamefont
  {Simon}}]{PhysRevB.88.075313}%
  \BibitemOpen
  \bibfield  {author} {\bibinfo {author} {\bibfnamefont {T.~S.}\ \bibnamefont
  {Jackson}}, \bibinfo {author} {\bibfnamefont {N.}~\bibnamefont {Read}}, \
  and\ \bibinfo {author} {\bibfnamefont {S.~H.}\ \bibnamefont {Simon}},\ }\href
  {\doibase 10.1103/PhysRevB.88.075313} {\bibfield  {journal} {\bibinfo
  {journal} {Phys. Rev. B}\ }\textbf {\bibinfo {volume} {88}},\ \bibinfo
  {pages} {075313} (\bibinfo {year} {2013})}\BibitemShut {NoStop}%
\bibitem [{\citenamefont {Thomale}\ \emph
  {et~al.}(2010{\natexlab{b}})\citenamefont {Thomale}, \citenamefont {Arovas},\
  and\ \citenamefont {Bernevig}}]{PhysRevLett.105.116805}%
  \BibitemOpen
  \bibfield  {author} {\bibinfo {author} {\bibfnamefont {R.}~\bibnamefont
  {Thomale}}, \bibinfo {author} {\bibfnamefont {D.~P.}\ \bibnamefont {Arovas}},
  \ and\ \bibinfo {author} {\bibfnamefont {B.~A.}\ \bibnamefont {Bernevig}},\
  }\href {\doibase 10.1103/PhysRevLett.105.116805} {\bibfield  {journal}
  {\bibinfo  {journal} {Phys. Rev. Lett.}\ }\textbf {\bibinfo {volume} {105}},\
  \bibinfo {pages} {116805} (\bibinfo {year} {2010}{\natexlab{b}})}\BibitemShut
  {NoStop}%
\bibitem [{\citenamefont {{Alba}}\ \emph {et~al.}(2012)\citenamefont {{Alba}},
  \citenamefont {{Haque}},\ and\ \citenamefont
  {{L{\"a}uchli}}}]{2012JSMTE..08..011A}%
  \BibitemOpen
  \bibfield  {author} {\bibinfo {author} {\bibfnamefont {V.}~\bibnamefont
  {{Alba}}}, \bibinfo {author} {\bibfnamefont {M.}~\bibnamefont {{Haque}}}, \
  and\ \bibinfo {author} {\bibfnamefont {A.~M.}\ \bibnamefont
  {{L{\"a}uchli}}},\ }\href {\doibase 10.1088/1742-5468/2012/08/P08011}
  {\bibfield  {journal} {\bibinfo  {journal} {J. Stat. Mech. Theor. Exp.}\
  }\textbf {\bibinfo {volume} {8}},\ \bibinfo {pages} {11} (\bibinfo {year}
  {2012})}\BibitemShut {NoStop}%
\bibitem [{\citenamefont {Pollmann}\ \emph {et~al.}(2010)\citenamefont
  {Pollmann}, \citenamefont {Turner}, \citenamefont {Berg},\ and\ \citenamefont
  {Oshikawa}}]{Pollmann:prb10}%
  \BibitemOpen
  \bibfield  {author} {\bibinfo {author} {\bibfnamefont {F.}~\bibnamefont
  {Pollmann}}, \bibinfo {author} {\bibfnamefont {A.~M.}\ \bibnamefont
  {Turner}}, \bibinfo {author} {\bibfnamefont {E.}~\bibnamefont {Berg}}, \ and\
  \bibinfo {author} {\bibfnamefont {M.}~\bibnamefont {Oshikawa}},\ }\href
  {\doibase 10.1103/PhysRevB.81.064439} {\bibfield  {journal} {\bibinfo
  {journal} {Phys. Rev. B}\ }\textbf {\bibinfo {volume} {81}},\ \bibinfo
  {pages} {064439} (\bibinfo {year} {2010})}\BibitemShut {NoStop}%
\bibitem [{\citenamefont {De~Chiara}\ \emph {et~al.}(2012)\citenamefont
  {De~Chiara}, \citenamefont {Lepori}, \citenamefont {Lewenstein},\ and\
  \citenamefont {Sanpera}}]{PhysRevLett.109.237208}%
  \BibitemOpen
  \bibfield  {author} {\bibinfo {author} {\bibfnamefont {G.}~\bibnamefont
  {De~Chiara}}, \bibinfo {author} {\bibfnamefont {L.}~\bibnamefont {Lepori}},
  \bibinfo {author} {\bibfnamefont {M.}~\bibnamefont {Lewenstein}}, \ and\
  \bibinfo {author} {\bibfnamefont {A.}~\bibnamefont {Sanpera}},\ }\href
  {\doibase 10.1103/PhysRevLett.109.237208} {\bibfield  {journal} {\bibinfo
  {journal} {Phys. Rev. Lett.}\ }\textbf {\bibinfo {volume} {109}},\ \bibinfo
  {pages} {237208} (\bibinfo {year} {2012})}\BibitemShut {NoStop}%
\bibitem [{\citenamefont {Ramkarthik}\ \emph {et~al.}(2013)\citenamefont
  {Ramkarthik}, \citenamefont {Chandra},\ and\ \citenamefont
  {Lakshminarayan}}]{PhysRevA.87.012302}%
  \BibitemOpen
  \bibfield  {author} {\bibinfo {author} {\bibfnamefont {M.~S.}\ \bibnamefont
  {Ramkarthik}}, \bibinfo {author} {\bibfnamefont {V.~R.}\ \bibnamefont
  {Chandra}}, \ and\ \bibinfo {author} {\bibfnamefont {A.}~\bibnamefont
  {Lakshminarayan}},\ }\href {\doibase 10.1103/PhysRevA.87.012302} {\bibfield
  {journal} {\bibinfo  {journal} {Phys. Rev. A}\ }\textbf {\bibinfo {volume}
  {87}},\ \bibinfo {pages} {012302} (\bibinfo {year} {2013})}\BibitemShut
  {NoStop}%
\bibitem [{\citenamefont {Franchini}\ \emph {et~al.}(2011)\citenamefont
  {Franchini}, \citenamefont {Its}, \citenamefont {Korepin},\ and\
  \citenamefont {Takhtajan}}]{Franchini:2010kq}%
  \BibitemOpen
  \bibfield  {author} {\bibinfo {author} {\bibfnamefont {F.}~\bibnamefont
  {Franchini}}, \bibinfo {author} {\bibfnamefont {A.}~\bibnamefont {Its}},
  \bibinfo {author} {\bibfnamefont {V.}~\bibnamefont {Korepin}}, \ and\
  \bibinfo {author} {\bibfnamefont {L.}~\bibnamefont {Takhtajan}},\ }\href@noop
  {} {\bibfield  {journal} {\bibinfo  {journal} {Quant.Inf.Proc.}\ }\textbf
  {\bibinfo {volume} {10}},\ \bibinfo {pages} {325} (\bibinfo {year}
  {2011})}\BibitemShut {NoStop}%
\bibitem [{\citenamefont {Lepori}\ \emph {et~al.}(2013)\citenamefont {Lepori},
  \citenamefont {De~Chiara},\ and\ \citenamefont
  {Sanpera}}]{PhysRevB.87.235107}%
  \BibitemOpen
  \bibfield  {author} {\bibinfo {author} {\bibfnamefont {L.}~\bibnamefont
  {Lepori}}, \bibinfo {author} {\bibfnamefont {G.}~\bibnamefont {De~Chiara}}, \
  and\ \bibinfo {author} {\bibfnamefont {A.}~\bibnamefont {Sanpera}},\ }\href
  {\doibase 10.1103/PhysRevB.87.235107} {\bibfield  {journal} {\bibinfo
  {journal} {Phys. Rev. B}\ }\textbf {\bibinfo {volume} {87}},\ \bibinfo
  {pages} {235107} (\bibinfo {year} {2013})}\BibitemShut {NoStop}%
\bibitem [{\citenamefont {Poilblanc}(2010)}]{Poilblanc:prl10}%
  \BibitemOpen
  \bibfield  {author} {\bibinfo {author} {\bibfnamefont {D.}~\bibnamefont
  {Poilblanc}},\ }\href {\doibase 10.1103/PhysRevLett.105.077202} {\bibfield
  {journal} {\bibinfo  {journal} {Phys. Rev. Lett.}\ }\textbf {\bibinfo
  {volume} {105}},\ \bibinfo {pages} {077202} (\bibinfo {year}
  {2010})}\BibitemShut {NoStop}%
\bibitem [{\citenamefont {L\"auchli}\ and\ \citenamefont
  {Schliemann}(2012)}]{Lauchli:prb12}%
  \BibitemOpen
  \bibfield  {author} {\bibinfo {author} {\bibfnamefont {A.~M.}\ \bibnamefont
  {L\"auchli}}\ and\ \bibinfo {author} {\bibfnamefont {J.}~\bibnamefont
  {Schliemann}},\ }\href@noop {} {\bibfield  {journal} {\bibinfo  {journal}
  {Phys. Rev. B}\ }\textbf {\bibinfo {volume} {85}},\ \bibinfo {pages} {054403}
  (\bibinfo {year} {2012})}\BibitemShut {NoStop}%
\bibitem [{\citenamefont {Lundgren}\ \emph {et~al.}(2012)\citenamefont
  {Lundgren}, \citenamefont {Chua},\ and\ \citenamefont {Fiete}}]{Lundgren}%
  \BibitemOpen
  \bibfield  {author} {\bibinfo {author} {\bibfnamefont {R.}~\bibnamefont
  {Lundgren}}, \bibinfo {author} {\bibfnamefont {V.}~\bibnamefont {Chua}}, \
  and\ \bibinfo {author} {\bibfnamefont {G.~A.}\ \bibnamefont {Fiete}},\ }\href
  {\doibase 10.1103/PhysRevB.86.224422} {\bibfield  {journal} {\bibinfo
  {journal} {Phys. Rev. B}\ }\textbf {\bibinfo {volume} {86}},\ \bibinfo
  {pages} {224422} (\bibinfo {year} {2012})}\BibitemShut {NoStop}%
\bibitem [{\citenamefont {Chen}\ and\ \citenamefont
  {Fradkin}(2013)}]{Fradkin_Ladder}%
  \BibitemOpen
  \bibfield  {author} {\bibinfo {author} {\bibfnamefont {X.}~\bibnamefont
  {Chen}}\ and\ \bibinfo {author} {\bibfnamefont {E.}~\bibnamefont {Fradkin}},\
  }\href {http://stacks.iop.org/1742-5468/2013/i=08/a=P08013} {\bibfield
  {journal} {\bibinfo  {journal} {J. Stat. Mech. Theor. Exp.}\ }\textbf
  {\bibinfo {volume} {2013}},\ \bibinfo {pages} {P08013} (\bibinfo {year}
  {2013})}\BibitemShut {NoStop}%
\bibitem [{\citenamefont {Cirac}\ \emph {et~al.}(2011)\citenamefont {Cirac},
  \citenamefont {Poilblanc}, \citenamefont {Schuch},\ and\ \citenamefont
  {Verstraete}}]{Cirac:prb11}%
  \BibitemOpen
  \bibfield  {author} {\bibinfo {author} {\bibfnamefont {J.~I.}\ \bibnamefont
  {Cirac}}, \bibinfo {author} {\bibfnamefont {D.}~\bibnamefont {Poilblanc}},
  \bibinfo {author} {\bibfnamefont {N.}~\bibnamefont {Schuch}}, \ and\ \bibinfo
  {author} {\bibfnamefont {F.}~\bibnamefont {Verstraete}},\ }\href {\doibase
  10.1103/PhysRevB.83.245134} {\bibfield  {journal} {\bibinfo  {journal} {Phys.
  Rev. B}\ }\textbf {\bibinfo {volume} {83}},\ \bibinfo {pages} {245134}
  (\bibinfo {year} {2011})}\BibitemShut {NoStop}%
\bibitem [{\citenamefont {Lundgren}\ \emph {et~al.}(2013)\citenamefont
  {Lundgren}, \citenamefont {Fuji}, \citenamefont {Furukawa},\ and\
  \citenamefont {Oshikawa}}]{PhysRevB.88.245137}%
  \BibitemOpen
  \bibfield  {author} {\bibinfo {author} {\bibfnamefont {R.}~\bibnamefont
  {Lundgren}}, \bibinfo {author} {\bibfnamefont {Y.}~\bibnamefont {Fuji}},
  \bibinfo {author} {\bibfnamefont {S.}~\bibnamefont {Furukawa}}, \ and\
  \bibinfo {author} {\bibfnamefont {M.}~\bibnamefont {Oshikawa}},\ }\href
  {\doibase 10.1103/PhysRevB.88.245137} {\bibfield  {journal} {\bibinfo
  {journal} {Phys. Rev. B}\ }\textbf {\bibinfo {volume} {88}},\ \bibinfo
  {pages} {245137} (\bibinfo {year} {2013})}\BibitemShut {NoStop}%
\bibitem [{\citenamefont {Tanaka}\ \emph {et~al.}(2012)\citenamefont {Tanaka},
  \citenamefont {Tamura},\ and\ \citenamefont {Katsura}}]{Tanaka:pra12}%
  \BibitemOpen
  \bibfield  {author} {\bibinfo {author} {\bibfnamefont {S.}~\bibnamefont
  {Tanaka}}, \bibinfo {author} {\bibfnamefont {R.}~\bibnamefont {Tamura}}, \
  and\ \bibinfo {author} {\bibfnamefont {H.}~\bibnamefont {Katsura}},\ }\href
  {\doibase 10.1103/PhysRevA.86.032326} {\bibfield  {journal} {\bibinfo
  {journal} {Phys. Rev. A}\ }\textbf {\bibinfo {volume} {86}},\ \bibinfo
  {pages} {032326} (\bibinfo {year} {2012})}\BibitemShut {NoStop}%
\bibitem [{\citenamefont {Alba}\ \emph {et~al.}(2012)\citenamefont {Alba},
  \citenamefont {Haque},\ and\ \citenamefont
  {L\"auchli}}]{PhysRevLett.108.227201}%
  \BibitemOpen
  \bibfield  {author} {\bibinfo {author} {\bibfnamefont {V.}~\bibnamefont
  {Alba}}, \bibinfo {author} {\bibfnamefont {M.}~\bibnamefont {Haque}}, \ and\
  \bibinfo {author} {\bibfnamefont {A.~M.}\ \bibnamefont {L\"auchli}},\ }\href
  {\doibase 10.1103/PhysRevLett.108.227201} {\bibfield  {journal} {\bibinfo
  {journal} {Phys. Rev. Lett.}\ }\textbf {\bibinfo {volume} {108}},\ \bibinfo
  {pages} {227201} (\bibinfo {year} {2012})}\BibitemShut {NoStop}%
\bibitem [{\citenamefont {{You}}\ \emph {et~al.}(2012)\citenamefont {{You}},
  \citenamefont {{Ole{\'s}}},\ and\ \citenamefont
  {{Horsch}}}]{2012PhRvB..86i4412Y}%
  \BibitemOpen
  \bibfield  {author} {\bibinfo {author} {\bibfnamefont {W.-L.}\ \bibnamefont
  {{You}}}, \bibinfo {author} {\bibfnamefont {A.~M.}\ \bibnamefont
  {{Ole{\'s}}}}, \ and\ \bibinfo {author} {\bibfnamefont {P.}~\bibnamefont
  {{Horsch}}},\ }\href {\doibase 10.1103/PhysRevB.86.094412} {\bibfield
  {journal} {\bibinfo  {journal} {\prb}\ }\textbf {\bibinfo {volume} {86}},\
  \bibinfo {eid} {094412} (\bibinfo {year} {2012})},\ \Eprint
  {http://arxiv.org/abs/1206.1062} {arXiv:1206.1062 [cond-mat.str-el]}
  \BibitemShut {NoStop}%
\bibitem [{\citenamefont {Kargarian}\ and\ \citenamefont
  {Fiete}(2010)}]{Kargarian:prb10}%
  \BibitemOpen
  \bibfield  {author} {\bibinfo {author} {\bibfnamefont {M.}~\bibnamefont
  {Kargarian}}\ and\ \bibinfo {author} {\bibfnamefont {G.~A.}\ \bibnamefont
  {Fiete}},\ }\href {\doibase 10.1103/PhysRevB.82.085106} {\bibfield  {journal}
  {\bibinfo  {journal} {Phys. Rev. B}\ }\textbf {\bibinfo {volume} {82}},\
  \bibinfo {pages} {085106} (\bibinfo {year} {2010})}\BibitemShut {NoStop}%
\bibitem [{\citenamefont {Fidkowski}(2010)}]{PhysRevLett.104.130502}%
  \BibitemOpen
  \bibfield  {author} {\bibinfo {author} {\bibfnamefont {L.}~\bibnamefont
  {Fidkowski}},\ }\href {\doibase 10.1103/PhysRevLett.104.130502} {\bibfield
  {journal} {\bibinfo  {journal} {Phys. Rev. Lett.}\ }\textbf {\bibinfo
  {volume} {104}},\ \bibinfo {pages} {130502} (\bibinfo {year}
  {2010})}\BibitemShut {NoStop}%
\bibitem [{\citenamefont {Turner}\ \emph {et~al.}(2010)\citenamefont {Turner},
  \citenamefont {Zhang},\ and\ \citenamefont
  {Vishwanath}}]{PhysRevB.82.241102}%
  \BibitemOpen
  \bibfield  {author} {\bibinfo {author} {\bibfnamefont {A.~M.}\ \bibnamefont
  {Turner}}, \bibinfo {author} {\bibfnamefont {Y.}~\bibnamefont {Zhang}}, \
  and\ \bibinfo {author} {\bibfnamefont {A.}~\bibnamefont {Vishwanath}},\
  }\href {\doibase 10.1103/PhysRevB.82.241102} {\bibfield  {journal} {\bibinfo
  {journal} {Phys. Rev. B}\ }\textbf {\bibinfo {volume} {82}},\ \bibinfo
  {pages} {241102} (\bibinfo {year} {2010})}\BibitemShut {NoStop}%
\bibitem [{\citenamefont {Fang}\ \emph {et~al.}(2013)\citenamefont {Fang},
  \citenamefont {Gilbert},\ and\ \citenamefont
  {Bernevig}}]{PhysRevB.87.035119}%
  \BibitemOpen
  \bibfield  {author} {\bibinfo {author} {\bibfnamefont {C.}~\bibnamefont
  {Fang}}, \bibinfo {author} {\bibfnamefont {M.~J.}\ \bibnamefont {Gilbert}}, \
  and\ \bibinfo {author} {\bibfnamefont {B.~A.}\ \bibnamefont {Bernevig}},\
  }\href {\doibase 10.1103/PhysRevB.87.035119} {\bibfield  {journal} {\bibinfo
  {journal} {Phys. Rev. B}\ }\textbf {\bibinfo {volume} {87}},\ \bibinfo
  {pages} {035119} (\bibinfo {year} {2013})}\BibitemShut {NoStop}%
\bibitem [{\citenamefont {Flammia}\ \emph {et~al.}(2009)\citenamefont
  {Flammia}, \citenamefont {Hamma}, \citenamefont {Hughes},\ and\ \citenamefont
  {Wen}}]{PhysRevLett.103.261601}%
  \BibitemOpen
  \bibfield  {author} {\bibinfo {author} {\bibfnamefont {S.~T.}\ \bibnamefont
  {Flammia}}, \bibinfo {author} {\bibfnamefont {A.}~\bibnamefont {Hamma}},
  \bibinfo {author} {\bibfnamefont {T.~L.}\ \bibnamefont {Hughes}}, \ and\
  \bibinfo {author} {\bibfnamefont {X.-G.}\ \bibnamefont {Wen}},\ }\href
  {\doibase 10.1103/PhysRevLett.103.261601} {\bibfield  {journal} {\bibinfo
  {journal} {Phys. Rev. Lett.}\ }\textbf {\bibinfo {volume} {103}},\ \bibinfo
  {pages} {261601} (\bibinfo {year} {2009})}\BibitemShut {NoStop}%
\bibitem [{\citenamefont {Alexandradinata}\ \emph {et~al.}(2011)\citenamefont
  {Alexandradinata}, \citenamefont {Hughes},\ and\ \citenamefont
  {Bernevig}}]{PhysRevB.84.195103}%
  \BibitemOpen
  \bibfield  {author} {\bibinfo {author} {\bibfnamefont {A.}~\bibnamefont
  {Alexandradinata}}, \bibinfo {author} {\bibfnamefont {T.~L.}\ \bibnamefont
  {Hughes}}, \ and\ \bibinfo {author} {\bibfnamefont {B.~A.}\ \bibnamefont
  {Bernevig}},\ }\href {\doibase 10.1103/PhysRevB.84.195103} {\bibfield
  {journal} {\bibinfo  {journal} {Phys. Rev. B}\ }\textbf {\bibinfo {volume}
  {84}},\ \bibinfo {pages} {195103} (\bibinfo {year} {2011})}\BibitemShut
  {NoStop}%
\bibitem [{\citenamefont {Kolley}\ \emph {et~al.}(2013)\citenamefont {Kolley},
  \citenamefont {Depenbrock}, \citenamefont {McCulloch}, \citenamefont
  {Schollw\"ock},\ and\ \citenamefont {Alba}}]{PhysRevB.88.144426}%
  \BibitemOpen
  \bibfield  {author} {\bibinfo {author} {\bibfnamefont {F.}~\bibnamefont
  {Kolley}}, \bibinfo {author} {\bibfnamefont {S.}~\bibnamefont {Depenbrock}},
  \bibinfo {author} {\bibfnamefont {I.~P.}\ \bibnamefont {McCulloch}}, \bibinfo
  {author} {\bibfnamefont {U.}~\bibnamefont {Schollw\"ock}}, \ and\ \bibinfo
  {author} {\bibfnamefont {V.}~\bibnamefont {Alba}},\ }\href {\doibase
  10.1103/PhysRevB.88.144426} {\bibfield  {journal} {\bibinfo  {journal} {Phys.
  Rev. B}\ }\textbf {\bibinfo {volume} {88}},\ \bibinfo {pages} {144426}
  (\bibinfo {year} {2013})}\BibitemShut {NoStop}%
\bibitem [{\citenamefont {{Alba}}\ \emph {et~al.}(2013)\citenamefont {{Alba}},
  \citenamefont {{Haque}},\ and\ \citenamefont
  {{L{\"a}uchli}}}]{2013PhRvL.110z0403A}%
  \BibitemOpen
  \bibfield  {author} {\bibinfo {author} {\bibfnamefont {V.}~\bibnamefont
  {{Alba}}}, \bibinfo {author} {\bibfnamefont {M.}~\bibnamefont {{Haque}}}, \
  and\ \bibinfo {author} {\bibfnamefont {A.~M.}\ \bibnamefont
  {{L{\"a}uchli}}},\ }\href {\doibase 10.1103/PhysRevLett.110.260403}
  {\bibfield  {journal} {\bibinfo  {journal} {Physical Review Letters}\
  }\textbf {\bibinfo {volume} {110}},\ \bibinfo {eid} {260403} (\bibinfo {year}
  {2013})}\BibitemShut {NoStop}%
\bibitem [{\citenamefont {{L{\"a}uchli}}(2013)}]{2013arXiv1303.0741L}%
  \BibitemOpen
  \bibfield  {author} {\bibinfo {author} {\bibfnamefont {A.~M.}\ \bibnamefont
  {{L{\"a}uchli}}},\ }\href@noop {} {\bibfield  {journal} {\bibinfo  {journal}
  {ArXiv e-prints}\ } (\bibinfo {year} {2013})},\ \Eprint
  {http://arxiv.org/abs/1303.0741} {arXiv:1303.0741 [cond-mat.stat-mech]}
  \BibitemShut {NoStop}%
\bibitem [{\citenamefont {Giampaolo}\ \emph {et~al.}(2013)\citenamefont
  {Giampaolo}, \citenamefont {Montangero}, \citenamefont {Dell'Anno},
  \citenamefont {De~Siena},\ and\ \citenamefont
  {Illuminati}}]{PhysRevB.88.125142}%
  \BibitemOpen
  \bibfield  {author} {\bibinfo {author} {\bibfnamefont {S.~M.}\ \bibnamefont
  {Giampaolo}}, \bibinfo {author} {\bibfnamefont {S.}~\bibnamefont
  {Montangero}}, \bibinfo {author} {\bibfnamefont {F.}~\bibnamefont
  {Dell'Anno}}, \bibinfo {author} {\bibfnamefont {S.}~\bibnamefont {De~Siena}},
  \ and\ \bibinfo {author} {\bibfnamefont {F.}~\bibnamefont {Illuminati}},\
  }\href {\doibase 10.1103/PhysRevB.88.125142} {\bibfield  {journal} {\bibinfo
  {journal} {Phys. Rev. B}\ }\textbf {\bibinfo {volume} {88}},\ \bibinfo
  {pages} {125142} (\bibinfo {year} {2013})}\BibitemShut {NoStop}%
\bibitem [{\citenamefont {Deng}\ and\ \citenamefont
  {Santos}(2011)}]{Deng:prb11}%
  \BibitemOpen
  \bibfield  {author} {\bibinfo {author} {\bibfnamefont {X.}~\bibnamefont
  {Deng}}\ and\ \bibinfo {author} {\bibfnamefont {L.}~\bibnamefont {Santos}},\
  }\href@noop {} {\bibfield  {journal} {\bibinfo  {journal} {Phys. Rev. B}\
  }\textbf {\bibinfo {volume} {84}},\ \bibinfo {pages} {085138} (\bibinfo
  {year} {2011})}\BibitemShut {NoStop}%
\bibitem [{\citenamefont {Turner}\ \emph {et~al.}(2011)\citenamefont {Turner},
  \citenamefont {Pollmann},\ and\ \citenamefont {Berg}}]{Turner:prb11}%
  \BibitemOpen
  \bibfield  {author} {\bibinfo {author} {\bibfnamefont {A.~M.}\ \bibnamefont
  {Turner}}, \bibinfo {author} {\bibfnamefont {F.}~\bibnamefont {Pollmann}}, \
  and\ \bibinfo {author} {\bibfnamefont {E.}~\bibnamefont {Berg}},\ }\href
  {\doibase 10.1103/PhysRevB.83.075102} {\bibfield  {journal} {\bibinfo
  {journal} {Phys. Rev. B}\ }\textbf {\bibinfo {volume} {83}},\ \bibinfo
  {pages} {075102} (\bibinfo {year} {2011})}\BibitemShut {NoStop}%
\bibitem [{\citenamefont {Pollmann}\ and\ \citenamefont
  {Moore}(2010)}]{Pollmann:njp10}%
  \BibitemOpen
  \bibfield  {author} {\bibinfo {author} {\bibfnamefont {F.}~\bibnamefont
  {Pollmann}}\ and\ \bibinfo {author} {\bibfnamefont {J.~E.}\ \bibnamefont
  {Moore}},\ }\href {http://stacks.iop.org/1367-2630/12/i=2/a=025006}
  {\bibfield  {journal} {\bibinfo  {journal} {New Journal of Physics}\ }\textbf
  {\bibinfo {volume} {12}},\ \bibinfo {pages} {025006} (\bibinfo {year}
  {2010})}\BibitemShut {NoStop}%
\bibitem [{\citenamefont {Hasebe}\ and\ \citenamefont
  {Totsuka}(2013)}]{PhysRevB.87.045115}%
  \BibitemOpen
  \bibfield  {author} {\bibinfo {author} {\bibfnamefont {K.}~\bibnamefont
  {Hasebe}}\ and\ \bibinfo {author} {\bibfnamefont {K.}~\bibnamefont
  {Totsuka}},\ }\href {\doibase 10.1103/PhysRevB.87.045115} {\bibfield
  {journal} {\bibinfo  {journal} {Phys. Rev. B}\ }\textbf {\bibinfo {volume}
  {87}},\ \bibinfo {pages} {045115} (\bibinfo {year} {2013})}\BibitemShut
  {NoStop}%
\bibitem [{\citenamefont {Kim}(2013)}]{PhysRevB.87.155120}%
  \BibitemOpen
  \bibfield  {author} {\bibinfo {author} {\bibfnamefont {I.~H.}\ \bibnamefont
  {Kim}},\ }\href {\doibase 10.1103/PhysRevB.87.155120} {\bibfield  {journal}
  {\bibinfo  {journal} {Phys. Rev. B}\ }\textbf {\bibinfo {volume} {87}},\
  \bibinfo {pages} {155120} (\bibinfo {year} {2013})}\BibitemShut {NoStop}%
\bibitem [{\citenamefont {Lou}\ \emph {et~al.}(2011)\citenamefont {Lou},
  \citenamefont {Tanaka}, \citenamefont {Katsura},\ and\ \citenamefont
  {Kawashima}}]{Lou:prb11}%
  \BibitemOpen
  \bibfield  {author} {\bibinfo {author} {\bibfnamefont {J.}~\bibnamefont
  {Lou}}, \bibinfo {author} {\bibfnamefont {S.}~\bibnamefont {Tanaka}},
  \bibinfo {author} {\bibfnamefont {H.}~\bibnamefont {Katsura}}, \ and\
  \bibinfo {author} {\bibfnamefont {N.}~\bibnamefont {Kawashima}},\ }\href
  {\doibase 10.1103/PhysRevB.84.245128} {\bibfield  {journal} {\bibinfo
  {journal} {Phys. Rev. B}\ }\textbf {\bibinfo {volume} {84}},\ \bibinfo
  {pages} {245128} (\bibinfo {year} {2011})}\BibitemShut {NoStop}%
\bibitem [{\citenamefont {Dubail}\ and\ \citenamefont
  {Read}(2011)}]{PhysRevLett.107.157001}%
  \BibitemOpen
  \bibfield  {author} {\bibinfo {author} {\bibfnamefont {J.}~\bibnamefont
  {Dubail}}\ and\ \bibinfo {author} {\bibfnamefont {N.}~\bibnamefont {Read}},\
  }\href {\doibase 10.1103/PhysRevLett.107.157001} {\bibfield  {journal}
  {\bibinfo  {journal} {Phys. Rev. Lett.}\ }\textbf {\bibinfo {volume} {107}},\
  \bibinfo {pages} {157001} (\bibinfo {year} {2011})}\BibitemShut {NoStop}%
\bibitem [{\citenamefont {Yao}\ and\ \citenamefont
  {Qi}(2010)}]{PhysRevLett.105.080501}%
  \BibitemOpen
  \bibfield  {author} {\bibinfo {author} {\bibfnamefont {H.}~\bibnamefont
  {Yao}}\ and\ \bibinfo {author} {\bibfnamefont {X.-L.}\ \bibnamefont {Qi}},\
  }\href {\doibase 10.1103/PhysRevLett.105.080501} {\bibfield  {journal}
  {\bibinfo  {journal} {Phys. Rev. Lett.}\ }\textbf {\bibinfo {volume} {105}},\
  \bibinfo {pages} {080501} (\bibinfo {year} {2010})}\BibitemShut {NoStop}%
\bibitem [{\citenamefont {James}\ and\ \citenamefont
  {Konik}(2013)}]{PhysRevB.87.241103}%
  \BibitemOpen
  \bibfield  {author} {\bibinfo {author} {\bibfnamefont {A.~J.~A.}\
  \bibnamefont {James}}\ and\ \bibinfo {author} {\bibfnamefont {R.~M.}\
  \bibnamefont {Konik}},\ }\href {\doibase 10.1103/PhysRevB.87.241103}
  {\bibfield  {journal} {\bibinfo  {journal} {Phys. Rev. B}\ }\textbf {\bibinfo
  {volume} {87}},\ \bibinfo {pages} {241103} (\bibinfo {year}
  {2013})}\BibitemShut {NoStop}%
\bibitem [{\citenamefont {Pouranvari}\ and\ \citenamefont
  {Yang}(2013)}]{PhysRevB.88.075123}%
  \BibitemOpen
  \bibfield  {author} {\bibinfo {author} {\bibfnamefont {M.}~\bibnamefont
  {Pouranvari}}\ and\ \bibinfo {author} {\bibfnamefont {K.}~\bibnamefont
  {Yang}},\ }\href {\doibase 10.1103/PhysRevB.88.075123} {\bibfield  {journal}
  {\bibinfo  {journal} {Phys. Rev. B}\ }\textbf {\bibinfo {volume} {88}},\
  \bibinfo {pages} {075123} (\bibinfo {year} {2013})}\BibitemShut {NoStop}%
\bibitem [{\citenamefont {Legner}\ and\ \citenamefont
  {Neupert}(2013)}]{PhysRevB.88.115114}%
  \BibitemOpen
  \bibfield  {author} {\bibinfo {author} {\bibfnamefont {M.}~\bibnamefont
  {Legner}}\ and\ \bibinfo {author} {\bibfnamefont {T.}~\bibnamefont
  {Neupert}},\ }\href {\doibase 10.1103/PhysRevB.88.115114} {\bibfield
  {journal} {\bibinfo  {journal} {Phys. Rev. B}\ }\textbf {\bibinfo {volume}
  {88}},\ \bibinfo {pages} {115114} (\bibinfo {year} {2013})}\BibitemShut
  {NoStop}%
\bibitem [{\citenamefont {Pi\ifmmode~\check{z}\else \v{z}\fi{}orn}\ \emph
  {et~al.}(2013)\citenamefont {Pi\ifmmode~\check{z}\else \v{z}\fi{}orn},
  \citenamefont {Verstraete},\ and\ \citenamefont
  {Konik}}]{PhysRevB.88.195102}%
  \BibitemOpen
  \bibfield  {author} {\bibinfo {author} {\bibfnamefont {I.}~\bibnamefont
  {Pi\ifmmode~\check{z}\else \v{z}\fi{}orn}}, \bibinfo {author} {\bibfnamefont
  {F.}~\bibnamefont {Verstraete}}, \ and\ \bibinfo {author} {\bibfnamefont
  {R.~M.}\ \bibnamefont {Konik}},\ }\href {\doibase 10.1103/PhysRevB.88.195102}
  {\bibfield  {journal} {\bibinfo  {journal} {Phys. Rev. B}\ }\textbf {\bibinfo
  {volume} {88}},\ \bibinfo {pages} {195102} (\bibinfo {year}
  {2013})}\BibitemShut {NoStop}%
\bibitem [{\citenamefont {Santos}(2013)}]{PhysRevB.87.035141}%
  \BibitemOpen
  \bibfield  {author} {\bibinfo {author} {\bibfnamefont {R.~A.}\ \bibnamefont
  {Santos}},\ }\href {\doibase 10.1103/PhysRevB.87.035141} {\bibfield
  {journal} {\bibinfo  {journal} {Phys. Rev. B}\ }\textbf {\bibinfo {volume}
  {87}},\ \bibinfo {pages} {035141} (\bibinfo {year} {2013})}\BibitemShut
  {NoStop}%
\bibitem [{\citenamefont {Regnault}\ and\ \citenamefont
  {Bernevig}(2011)}]{PhysRevX.1.021014}%
  \BibitemOpen
  \bibfield  {author} {\bibinfo {author} {\bibfnamefont {N.}~\bibnamefont
  {Regnault}}\ and\ \bibinfo {author} {\bibfnamefont {B.~A.}\ \bibnamefont
  {Bernevig}},\ }\href {\doibase 10.1103/PhysRevX.1.021014} {\bibfield
  {journal} {\bibinfo  {journal} {Phys. Rev. X}\ }\textbf {\bibinfo {volume}
  {1}},\ \bibinfo {pages} {021014} (\bibinfo {year} {2011})}\BibitemShut
  {NoStop}%
\bibitem [{\citenamefont {{Schliemann}}(2013)}]{2013NJPh...15e3017S}%
  \BibitemOpen
  \bibfield  {author} {\bibinfo {author} {\bibfnamefont {J.}~\bibnamefont
  {{Schliemann}}},\ }\href {\doibase 10.1088/1367-2630/15/5/053017} {\bibfield
  {journal} {\bibinfo  {journal} {New Journal of Physics}\ }\textbf {\bibinfo
  {volume} {15}},\ \bibinfo {eid} {053017} (\bibinfo {year}
  {2013})}\BibitemShut {NoStop}%
\bibitem [{\citenamefont {{Tubman}}\ and\ \citenamefont {{ChangMo
  Yang}}(2014)}]{2014arXiv1402.0503T}%
  \BibitemOpen
  \bibfield  {author} {\bibinfo {author} {\bibfnamefont {N.~M.}\ \bibnamefont
  {{Tubman}}}\ and\ \bibinfo {author} {\bibfnamefont {D.}~\bibnamefont
  {{ChangMo Yang}}},\ }\href@noop {} {\bibfield  {journal} {\bibinfo  {journal}
  {ArXiv e-prints}\ } (\bibinfo {year} {2014})},\ \Eprint
  {http://arxiv.org/abs/1402.0503} {arXiv:1402.0503 [cond-mat.str-el]}
  \BibitemShut {NoStop}%
\bibitem [{\citenamefont {Qi}\ \emph {et~al.}(2012)\citenamefont {Qi},
  \citenamefont {Katsura},\ and\ \citenamefont {Ludwig}}]{Qi:prl12}%
  \BibitemOpen
  \bibfield  {author} {\bibinfo {author} {\bibfnamefont {X.-L.}\ \bibnamefont
  {Qi}}, \bibinfo {author} {\bibfnamefont {H.}~\bibnamefont {Katsura}}, \ and\
  \bibinfo {author} {\bibfnamefont {A.~W.~W.}\ \bibnamefont {Ludwig}},\ }\href
  {\doibase 10.1103/PhysRevLett.108.196402} {\bibfield  {journal} {\bibinfo
  {journal} {Phys. Rev. Lett.}\ }\textbf {\bibinfo {volume} {108}},\ \bibinfo
  {pages} {196402} (\bibinfo {year} {2012})}\BibitemShut {NoStop}%
\bibitem [{\citenamefont {Chandran}\ \emph {et~al.}(2011)\citenamefont
  {Chandran}, \citenamefont {Hermanns}, \citenamefont {Regnault},\ and\
  \citenamefont {Bernevig}}]{PhysRevB.84.205136}%
  \BibitemOpen
  \bibfield  {author} {\bibinfo {author} {\bibfnamefont {A.}~\bibnamefont
  {Chandran}}, \bibinfo {author} {\bibfnamefont {M.}~\bibnamefont {Hermanns}},
  \bibinfo {author} {\bibfnamefont {N.}~\bibnamefont {Regnault}}, \ and\
  \bibinfo {author} {\bibfnamefont {B.~A.}\ \bibnamefont {Bernevig}},\ }\href
  {\doibase 10.1103/PhysRevB.84.205136} {\bibfield  {journal} {\bibinfo
  {journal} {Phys. Rev. B}\ }\textbf {\bibinfo {volume} {84}},\ \bibinfo
  {pages} {205136} (\bibinfo {year} {2011})}\BibitemShut {NoStop}%
\bibitem [{\citenamefont {Dubail}\ \emph
  {et~al.}(2012{\natexlab{b}})\citenamefont {Dubail}, \citenamefont {Read},\
  and\ \citenamefont {Rezayi}}]{Dubail_proof:prb12}%
  \BibitemOpen
  \bibfield  {author} {\bibinfo {author} {\bibfnamefont {J.}~\bibnamefont
  {Dubail}}, \bibinfo {author} {\bibfnamefont {N.}~\bibnamefont {Read}}, \ and\
  \bibinfo {author} {\bibfnamefont {E.~H.}\ \bibnamefont {Rezayi}},\ }\href
  {\doibase 10.1103/PhysRevB.86.245310} {\bibfield  {journal} {\bibinfo
  {journal} {Phys. Rev. B}\ }\textbf {\bibinfo {volume} {86}},\ \bibinfo
  {pages} {245310} (\bibinfo {year} {2012}{\natexlab{b}})}\BibitemShut
  {NoStop}%
\bibitem [{\citenamefont {Swingle}\ and\ \citenamefont
  {Senthil}(2012)}]{PhysRevB.86.045117}%
  \BibitemOpen
  \bibfield  {author} {\bibinfo {author} {\bibfnamefont {B.}~\bibnamefont
  {Swingle}}\ and\ \bibinfo {author} {\bibfnamefont {T.}~\bibnamefont
  {Senthil}},\ }\href {\doibase 10.1103/PhysRevB.86.045117} {\bibfield
  {journal} {\bibinfo  {journal} {Phys. Rev. B}\ }\textbf {\bibinfo {volume}
  {86}},\ \bibinfo {pages} {045117} (\bibinfo {year} {2012})}\BibitemShut
  {NoStop}%
\bibitem [{\citenamefont {Calabrese}\ and\ \citenamefont
  {Lefevre}(2008)}]{Calabrese:pra08}%
  \BibitemOpen
  \bibfield  {author} {\bibinfo {author} {\bibfnamefont {P.}~\bibnamefont
  {Calabrese}}\ and\ \bibinfo {author} {\bibfnamefont {A.}~\bibnamefont
  {Lefevre}},\ }\href {\doibase 10.1103/PhysRevA.78.032329} {\bibfield
  {journal} {\bibinfo  {journal} {Phys. Rev. A}\ }\textbf {\bibinfo {volume}
  {78}},\ \bibinfo {pages} {032329} (\bibinfo {year} {2008})}\BibitemShut
  {NoStop}%
\bibitem [{\citenamefont {Gogolin}\ \emph {et~al.}(1998)\citenamefont
  {Gogolin}, \citenamefont {Nersesyan},\ and\ \citenamefont
  {Tsvelik}}]{Gogolin:book}%
  \BibitemOpen
  \bibfield  {author} {\bibinfo {author} {\bibfnamefont {A.~O.}\ \bibnamefont
  {Gogolin}}, \bibinfo {author} {\bibfnamefont {A.~A.}\ \bibnamefont
  {Nersesyan}}, \ and\ \bibinfo {author} {\bibfnamefont {A.~M.}\ \bibnamefont
  {Tsvelik}},\ }\href@noop {} {\emph {\bibinfo {title} {Bosonization and
  Strongly Correlated Systems}}}\ (\bibinfo  {publisher} {Cambridge University
  Press, New York},\ \bibinfo {year} {1998})\BibitemShut {NoStop}%
\bibitem [{\citenamefont {Xiang}(1996)}]{PhysRevB.53.R10445}%
  \BibitemOpen
  \bibfield  {author} {\bibinfo {author} {\bibfnamefont {T.}~\bibnamefont
  {Xiang}},\ }\href {\doibase 10.1103/PhysRevB.53.R10445} {\bibfield  {journal}
  {\bibinfo  {journal} {Phys. Rev. B}\ }\textbf {\bibinfo {volume} {53}},\
  \bibinfo {pages} {R10445} (\bibinfo {year} {1996})}\BibitemShut {NoStop}%
\bibitem [{\citenamefont {Schollw\"ock}(2005)}]{RevModPhys.77.259}%
  \BibitemOpen
  \bibfield  {author} {\bibinfo {author} {\bibfnamefont {U.}~\bibnamefont
  {Schollw\"ock}},\ }\href {\doibase 10.1103/RevModPhys.77.259} {\bibfield
  {journal} {\bibinfo  {journal} {Rev. Mod. Phys.}\ }\textbf {\bibinfo {volume}
  {77}},\ \bibinfo {pages} {259} (\bibinfo {year} {2005})}\BibitemShut
  {NoStop}%
\bibitem [{\citenamefont {Hallberg}(2006)}]{hallberg}%
  \BibitemOpen
  \bibfield  {author} {\bibinfo {author} {\bibfnamefont {K.}~\bibnamefont
  {Hallberg}},\ }\href@noop {} {\bibfield  {journal} {\bibinfo  {journal} {Adv.
  Phys.}\ }\textbf {\bibinfo {volume} {55}},\ \bibinfo {pages} {477} (\bibinfo
  {year} {2006})}\BibitemShut {NoStop}%
\bibitem [{\citenamefont {Legeza}\ \emph {et~al.}(2008)\citenamefont {Legeza},
  \citenamefont {Noack}, \citenamefont {Solyom},\ and\ \citenamefont
  {Tincani}}]{legeza}%
  \BibitemOpen
  \bibfield  {author} {\bibinfo {author} {\bibfnamefont {O.}~\bibnamefont
  {Legeza}}, \bibinfo {author} {\bibfnamefont {R.~M.}\ \bibnamefont {Noack}},
  \bibinfo {author} {\bibfnamefont {J.}~\bibnamefont {Solyom}}, \ and\ \bibinfo
  {author} {\bibfnamefont {L.}~\bibnamefont {Tincani}},\ }\href@noop {} {\emph
  {\bibinfo {title} {Applications of Quantum Information in the Density-Matrix
  Renormalization Group}}},\ Vol.\ \bibinfo {volume} {739}\ (\bibinfo
  {publisher} {Springer},\ \bibinfo {address} {Berlin},\ \bibinfo {year}
  {2008})\BibitemShut {NoStop}%
\bibitem [{\citenamefont {F\"uhringer}\ \emph {et~al.}(2008)\citenamefont
  {F\"uhringer}, \citenamefont {Rachel}, \citenamefont {Thomale}, \citenamefont
  {Greiter},\ and\ \citenamefont {Schmitteckert}}]{sun}%
  \BibitemOpen
  \bibfield  {author} {\bibinfo {author} {\bibfnamefont {M.}~\bibnamefont
  {F\"uhringer}}, \bibinfo {author} {\bibfnamefont {S.}~\bibnamefont {Rachel}},
  \bibinfo {author} {\bibfnamefont {R.}~\bibnamefont {Thomale}}, \bibinfo
  {author} {\bibfnamefont {M.}~\bibnamefont {Greiter}}, \ and\ \bibinfo
  {author} {\bibfnamefont {P.}~\bibnamefont {Schmitteckert}},\ }\href@noop {}
  {\bibfield  {journal} {\bibinfo  {journal} {Ann. Phys. (Berlin)}\ }\textbf
  {\bibinfo {volume} {17}},\ \bibinfo {pages} {922} (\bibinfo {year}
  {2008})}\BibitemShut {NoStop}%
\bibitem [{\citenamefont {Mondragon-Shem}\ \emph {et~al.}(2013)\citenamefont
  {Mondragon-Shem}, \citenamefont {Khan},\ and\ \citenamefont
  {Hughes}}]{disorder_fermi_MS}%
  \BibitemOpen
  \bibfield  {author} {\bibinfo {author} {\bibfnamefont {I.}~\bibnamefont
  {Mondragon-Shem}}, \bibinfo {author} {\bibfnamefont {M.}~\bibnamefont
  {Khan}}, \ and\ \bibinfo {author} {\bibfnamefont {T.~L.}\ \bibnamefont
  {Hughes}},\ }\href {\doibase 10.1103/PhysRevLett.110.046806} {\bibfield
  {journal} {\bibinfo  {journal} {Phys. Rev. Lett.}\ }\textbf {\bibinfo
  {volume} {110}},\ \bibinfo {pages} {046806} (\bibinfo {year}
  {2013})}\BibitemShut {NoStop}%
\bibitem [{\citenamefont {{Mondragon-Shem}}\ and\ \citenamefont
  {{Hughes}}(2014)}]{2014arXiv1403.6129M}%
  \BibitemOpen
  \bibfield  {author} {\bibinfo {author} {\bibfnamefont {I.}~\bibnamefont
  {{Mondragon-Shem}}}\ and\ \bibinfo {author} {\bibfnamefont {T.~L.}\
  \bibnamefont {{Hughes}}},\ }\href@noop {} {\bibfield  {journal} {\bibinfo
  {journal} {ArXiv e-prints}\ } (\bibinfo {year} {2014})},\ \Eprint
  {http://arxiv.org/abs/1403.6129} {arXiv:1403.6129 [cond-mat.dis-nn]}
  \BibitemShut {NoStop}%
\bibitem [{\citenamefont {{Andrade}}\ \emph {et~al.}(2014)\citenamefont
  {{Andrade}}, \citenamefont {{Steudtner}},\ and\ \citenamefont
  {{Vojta}}}]{2014arXiv1403.2599A}%
  \BibitemOpen
  \bibfield  {author} {\bibinfo {author} {\bibfnamefont {E.~C.}\ \bibnamefont
  {{Andrade}}}, \bibinfo {author} {\bibfnamefont {M.}~\bibnamefont
  {{Steudtner}}}, \ and\ \bibinfo {author} {\bibfnamefont {M.}~\bibnamefont
  {{Vojta}}},\ }\href@noop {} {\bibfield  {journal} {\bibinfo  {journal} {ArXiv
  e-prints}\ } (\bibinfo {year} {2014})},\ \Eprint
  {http://arxiv.org/abs/1403.2599} {arXiv:1403.2599 [cond-mat.dis-nn]}
  \BibitemShut {NoStop}%
\bibitem [{\citenamefont {Balasubramanian}\ \emph {et~al.}(2012)\citenamefont
  {Balasubramanian}, \citenamefont {McDermott},\ and\ \citenamefont
  {Van~Raamsdonk}}]{PhysRevD.86.045014}%
  \BibitemOpen
  \bibfield  {author} {\bibinfo {author} {\bibfnamefont {V.}~\bibnamefont
  {Balasubramanian}}, \bibinfo {author} {\bibfnamefont {M.~B.}\ \bibnamefont
  {McDermott}}, \ and\ \bibinfo {author} {\bibfnamefont {M.}~\bibnamefont
  {Van~Raamsdonk}},\ }\href {\doibase 10.1103/PhysRevD.86.045014} {\bibfield
  {journal} {\bibinfo  {journal} {Phys. Rev. D}\ }\textbf {\bibinfo {volume}
  {86}},\ \bibinfo {pages} {045014} (\bibinfo {year} {2012})}\BibitemShut
  {NoStop}%
\bibitem [{\citenamefont {Haldane}(1988)}]{PhysRevLett.60.635}%
  \BibitemOpen
  \bibfield  {author} {\bibinfo {author} {\bibfnamefont {F.~D.~M.}\
  \bibnamefont {Haldane}},\ }\href {\doibase 10.1103/PhysRevLett.60.635}
  {\bibfield  {journal} {\bibinfo  {journal} {Phys. Rev. Lett.}\ }\textbf
  {\bibinfo {volume} {60}},\ \bibinfo {pages} {635} (\bibinfo {year}
  {1988})}\BibitemShut {NoStop}%
\bibitem [{\citenamefont {Shastry}(1988)}]{PhysRevLett.60.639}%
  \BibitemOpen
  \bibfield  {author} {\bibinfo {author} {\bibfnamefont {B.~S.}\ \bibnamefont
  {Shastry}},\ }\href {\doibase 10.1103/PhysRevLett.60.639} {\bibfield
  {journal} {\bibinfo  {journal} {Phys. Rev. Lett.}\ }\textbf {\bibinfo
  {volume} {60}},\ \bibinfo {pages} {639} (\bibinfo {year} {1988})}\BibitemShut
  {NoStop}%
\bibitem [{\citenamefont {Greiter}(2011)}]{Greiter}%
  \BibitemOpen
  \bibfield  {author} {\bibinfo {author} {\bibfnamefont {M.}~\bibnamefont
  {Greiter}},\ }\href@noop {} {\emph {\bibinfo {title} {Mapping of Parent
  Hamiltonians: from Abelian and non-Abelian Quantum Hall States to Exact
  Models of Critical Spin Chains}}}\ (\bibinfo  {publisher} {Springer Tract of
  Modern Physics},\ \bibinfo {year} {2011})\BibitemShut {NoStop}%
\bibitem [{\citenamefont {Thomale}\ \emph {et~al.}(2012)\citenamefont
  {Thomale}, \citenamefont {Rachel}, \citenamefont {Schmitteckert},\ and\
  \citenamefont {Greiter}}]{PhysRevB.85.195149}%
  \BibitemOpen
  \bibfield  {author} {\bibinfo {author} {\bibfnamefont {R.}~\bibnamefont
  {Thomale}}, \bibinfo {author} {\bibfnamefont {S.}~\bibnamefont {Rachel}},
  \bibinfo {author} {\bibfnamefont {P.}~\bibnamefont {Schmitteckert}}, \ and\
  \bibinfo {author} {\bibfnamefont {M.}~\bibnamefont {Greiter}},\ }\href
  {\doibase 10.1103/PhysRevB.85.195149} {\bibfield  {journal} {\bibinfo
  {journal} {Phys. Rev. B}\ }\textbf {\bibinfo {volume} {85}},\ \bibinfo
  {pages} {195149} (\bibinfo {year} {2012})}\BibitemShut {NoStop}%
\bibitem [{\citenamefont {{Chandran}}\ \emph {et~al.}(2013)\citenamefont
  {{Chandran}}, \citenamefont {{Khemani}},\ and\ \citenamefont
  {{Sondhi}}}]{2013arXiv1311.2946C}%
  \BibitemOpen
  \bibfield  {author} {\bibinfo {author} {\bibfnamefont {A.}~\bibnamefont
  {{Chandran}}}, \bibinfo {author} {\bibfnamefont {V.}~\bibnamefont
  {{Khemani}}}, \ and\ \bibinfo {author} {\bibfnamefont {S.~L.}\ \bibnamefont
  {{Sondhi}}},\ }\href@noop {} {\bibfield  {journal} {\bibinfo  {journal}
  {ArXiv e-prints}\ } (\bibinfo {year} {2013})},\ \Eprint
  {http://arxiv.org/abs/1311.2946} {arXiv:1311.2946 [cond-mat.str-el]}
  \BibitemShut {NoStop}%
\bibitem [{\citenamefont {{Braganca}}\ \emph {et~al.}(2013)\citenamefont
  {{Braganca}}, \citenamefont {{Mascarenhas}}, \citenamefont {{Luiz}},
  \citenamefont {{Duarte}}, \citenamefont {{Pereira}}, \citenamefont
  {{Santos}},\ and\ \citenamefont {{Aguiar}}}]{2013arXiv1312.0619B}%
  \BibitemOpen
  \bibfield  {author} {\bibinfo {author} {\bibfnamefont {H.}~\bibnamefont
  {{Braganca}}}, \bibinfo {author} {\bibfnamefont {E.}~\bibnamefont
  {{Mascarenhas}}}, \bibinfo {author} {\bibfnamefont {G.~I.}\ \bibnamefont
  {{Luiz}}}, \bibinfo {author} {\bibfnamefont {C.}~\bibnamefont {{Duarte}}},
  \bibinfo {author} {\bibfnamefont {R.~G.}\ \bibnamefont {{Pereira}}}, \bibinfo
  {author} {\bibfnamefont {M.~F.}\ \bibnamefont {{Santos}}}, \ and\ \bibinfo
  {author} {\bibfnamefont {M.~C.~O.}\ \bibnamefont {{Aguiar}}},\ }\href@noop {}
  {\bibfield  {journal} {\bibinfo  {journal} {ArXiv e-prints}\ } (\bibinfo
  {year} {2013})},\ \Eprint {http://arxiv.org/abs/1312.0619} {arXiv:1312.0619
  [cond-mat.str-el]} \BibitemShut {NoStop}%
\bibitem [{\citenamefont {Levin}\ and\ \citenamefont
  {Wen}(2006)}]{PhysRevLett.96.110405}%
  \BibitemOpen
  \bibfield  {author} {\bibinfo {author} {\bibfnamefont {M.}~\bibnamefont
  {Levin}}\ and\ \bibinfo {author} {\bibfnamefont {X.-G.}\ \bibnamefont
  {Wen}},\ }\href {\doibase 10.1103/PhysRevLett.96.110405} {\bibfield
  {journal} {\bibinfo  {journal} {Phys. Rev. Lett.}\ }\textbf {\bibinfo
  {volume} {96}},\ \bibinfo {pages} {110405} (\bibinfo {year}
  {2006})}\BibitemShut {NoStop}%
\bibitem [{\citenamefont {Kitaev}\ and\ \citenamefont
  {Preskill}(2006)}]{PhysRevLett.96.110404}%
  \BibitemOpen
  \bibfield  {author} {\bibinfo {author} {\bibfnamefont {A.}~\bibnamefont
  {Kitaev}}\ and\ \bibinfo {author} {\bibfnamefont {J.}~\bibnamefont
  {Preskill}},\ }\href {\doibase 10.1103/PhysRevLett.96.110404} {\bibfield
  {journal} {\bibinfo  {journal} {Phys. Rev. Lett.}\ }\textbf {\bibinfo
  {volume} {96}},\ \bibinfo {pages} {110404} (\bibinfo {year}
  {2006})}\BibitemShut {NoStop}%
\bibitem [{\citenamefont {Haque}\ \emph {et~al.}(2007)\citenamefont {Haque},
  \citenamefont {Zozulya},\ and\ \citenamefont
  {Schoutens}}]{PhysRevLett.98.060401}%
  \BibitemOpen
  \bibfield  {author} {\bibinfo {author} {\bibfnamefont {M.}~\bibnamefont
  {Haque}}, \bibinfo {author} {\bibfnamefont {O.}~\bibnamefont {Zozulya}}, \
  and\ \bibinfo {author} {\bibfnamefont {K.}~\bibnamefont {Schoutens}},\ }\href
  {\doibase 10.1103/PhysRevLett.98.060401} {\bibfield  {journal} {\bibinfo
  {journal} {Phys. Rev. Lett.}\ }\textbf {\bibinfo {volume} {98}},\ \bibinfo
  {pages} {060401} (\bibinfo {year} {2007})}\BibitemShut {NoStop}%
\bibitem [{Sup()}]{Supplemental}%
  \BibitemOpen
  \href@noop {} {}\bibinfo {note} {See Supplemental Material}\BibitemShut {NoStop}%
\bibitem [{\citenamefont {Peschel}(2003)}]{2003JPhA...36L.205P}
\BibitemOpen
  \bibfield  {author} {\bibinfo {author} {\bibfnamefont {I.}~\bibnamefont
  {Peschel}},\ }\href {\doibase 10.1088/0305-4470/36/14/101} {\bibfield
  {journal} {\bibinfo  {journal} {J. Physics A: Mathematical and General}\
  }\textbf {\bibinfo {volume} {36}},\ \bibinfo {pages} {L205} (\bibinfo {year}
  {2003})}\BibitemShut {NoStop}%
\bibitem [{\citenamefont {Giamarchi}(2013)}]{Giamarchi:book}%
  \BibitemOpen
  \bibfield  {author} {\bibinfo {author} {\bibfnamefont {T.}~\bibnamefont
  {Giamarchi}},\ }\href@noop {} {\emph {\bibinfo {title} {Quantum Physics in
  One Dimension}}}\ (\bibinfo  {publisher} {Oxford University Press, New
  York},\ \bibinfo {year} {2013})\BibitemShut {NoStop}%
\bibitem [{\citenamefont {Furukawa}\ \emph {et~al.}(2012)\citenamefont
  {Furukawa}, \citenamefont {Sato}, \citenamefont {Onoda},\ and\ \citenamefont
  {Furusaki}}]{PhysRevB.86.094417}%
  \BibitemOpen
  \bibfield  {author} {\bibinfo {author} {\bibfnamefont {S.}~\bibnamefont
  {Furukawa}}, \bibinfo {author} {\bibfnamefont {M.}~\bibnamefont {Sato}},
  \bibinfo {author} {\bibfnamefont {S.}~\bibnamefont {Onoda}}, \ and\ \bibinfo
  {author} {\bibfnamefont {A.}~\bibnamefont {Furusaki}},\ }\href {\doibase
  10.1103/PhysRevB.86.094417} {\bibfield  {journal} {\bibinfo  {journal} {Phys.
  Rev. B}\ }\textbf {\bibinfo {volume} {86}},\ \bibinfo {pages} {094417}
  (\bibinfo {year} {2012})}\BibitemShut {NoStop}%
\bibitem [{\citenamefont {Eisert}\ \emph {et~al.}(2010)\citenamefont {Eisert},
  \citenamefont {Cramer},\ and\ \citenamefont {Plenio}}]{RevModPhys.82.277}%
  \BibitemOpen
  \bibfield  {author} {\bibinfo {author} {\bibfnamefont {J.}~\bibnamefont
  {Eisert}}, \bibinfo {author} {\bibfnamefont {M.}~\bibnamefont {Cramer}}, \
  and\ \bibinfo {author} {\bibfnamefont {M.~B.}\ \bibnamefont {Plenio}},\
  }\href {\doibase 10.1103/RevModPhys.82.277} {\bibfield  {journal} {\bibinfo
  {journal} {Rev. Mod. Phys.}\ }\textbf {\bibinfo {volume} {82}},\ \bibinfo
  {pages} {277} (\bibinfo {year} {2010})}\BibitemShut {NoStop}%
\bibitem [{\citenamefont {Benatti}\ \emph {et~al.}(2014)\citenamefont
  {Benatti}, \citenamefont {Floreanini},\ and\ \citenamefont
  {Marzolino}}]{PhysRevA.89.032326}%
  \BibitemOpen
  \bibfield  {author} {\bibinfo {author} {\bibfnamefont {F.}~\bibnamefont
  {Benatti}}, \bibinfo {author} {\bibfnamefont {R.}~\bibnamefont {Floreanini}},
  \ and\ \bibinfo {author} {\bibfnamefont {U.}~\bibnamefont {Marzolino}},\
  }\href {\doibase 10.1103/PhysRevA.89.032326} {\bibfield  {journal} {\bibinfo
  {journal} {Phys. Rev. A}\ }\textbf {\bibinfo {volume} {89}},\ \bibinfo
  {pages} {032326} (\bibinfo {year} {2014})}\BibitemShut {NoStop}%
\bibitem [{\citenamefont {{Benatti}}\ \emph {et~al.}(2010)\citenamefont
  {{Benatti}}, \citenamefont {{Floreanini}},\ and\ \citenamefont
  {{Marzolino}}}]{2010AnPhy.325..924B}%
  \BibitemOpen
  \bibfield  {author} {\bibinfo {author} {\bibfnamefont {F.}~\bibnamefont
  {{Benatti}}}, \bibinfo {author} {\bibfnamefont {R.}~\bibnamefont
  {{Floreanini}}}, \ and\ \bibinfo {author} {\bibfnamefont {U.}~\bibnamefont
  {{Marzolino}}},\ }\href {\doibase 10.1016/j.aop.2010.01.005} {\bibfield
  {journal} {\bibinfo  {journal} {Annals of Physics}\ }\textbf {\bibinfo
  {volume} {325}},\ \bibinfo {pages} {924} (\bibinfo {year}
  {2010})}\BibitemShut {NoStop}%
\bibitem [{\citenamefont {Gu}\ \emph {et~al.}(2002)\citenamefont {Gu},
  \citenamefont {Pereira},\ and\ \citenamefont {Peres}}]{PhysRevB.66.235108}%
  \BibitemOpen
  \bibfield  {author} {\bibinfo {author} {\bibfnamefont {S.-J.}\ \bibnamefont
  {Gu}}, \bibinfo {author} {\bibfnamefont {V.~M.}\ \bibnamefont {Pereira}}, \
  and\ \bibinfo {author} {\bibfnamefont {N.~M.~R.}\ \bibnamefont {Peres}},\
  }\href {\doibase 10.1103/PhysRevB.66.235108} {\bibfield  {journal} {\bibinfo
  {journal} {Phys. Rev. B}\ }\textbf {\bibinfo {volume} {66}},\ \bibinfo
  {pages} {235108} (\bibinfo {year} {2002})}\BibitemShut {NoStop}%
\bibitem [{\citenamefont {Bonner}\ and\ \citenamefont
  {Fisher}(1964)}]{PhysRev.135.A640}%
  \BibitemOpen
  \bibfield  {author} {\bibinfo {author} {\bibfnamefont {J.~C.}\ \bibnamefont
  {Bonner}}\ and\ \bibinfo {author} {\bibfnamefont {M.~E.}\ \bibnamefont
  {Fisher}},\ }\href {\doibase 10.1103/PhysRev.135.A640} {\bibfield  {journal}
  {\bibinfo  {journal} {Phys. Rev.}\ }\textbf {\bibinfo {volume} {135}},\
  \bibinfo {pages} {A640} (\bibinfo {year} {1964})}\BibitemShut {NoStop}%
\end{thebibliography}
\end{document}